\documentclass[11pt]{elsarticle}
\usepackage{lineno,hyperref}
\modulolinenumbers[5]

\journal{Journal of \LaTeX\ Templates}
\usepackage{float}
\usepackage{enumerate}
\usepackage{tikz,pgffor}

\usetikzlibrary{shapes}

\usetikzlibrary{arrows}
\usepackage{epstopdf}
\usepackage{inputenc}
\usepackage{epsfig}
\usepackage{amsmath}
\usepackage{amsfonts}
\usepackage{latexsym}
\usepackage{color}

\newtheorem{lemma}{Lemma}[section]

\newtheorem{prop}{Proposition}[section]
\newtheorem{definition}{Definition}

\newtheorem{Cor}{Corollary}[section]

\numberwithin{equation}{section}
\numberwithin{figure}{section}

\def\proof
{\noindent
{\sc Proof.   }
}
\def\eop
{\hfill $\Box$ \\ }

\usepackage[text={6.5in,8.75in},centering,dvipdfm]{geometry}

\newcommand{\indep}{\;\, \rule[0em]{.03em}{.67em} \hspace{-.25em}
\rule[0em]{.65em}{.03em} \hspace{-.25em}
\rule[0em]{.03em}{.67em}\;\,}

\newcommand{\real}[1]{{\mathbb R}^{#1}}

\newcommand{\spn}{\mathrm{span}}

\newcommand{\E}{\mathrm{E}}
\newcommand{\var}{\mathrm{var}}
\newcommand{\cov}{\mathrm{cov}}
\newcommand{\cor}{\mathrm{cor}}

\newcommand{\rank}{\mathrm{rank}}

\newcommand{\tr}{\mathrm{\,tr}}

\newcommand{\Q}{{\mathbf Q}}

\newcommand{\tX}{\tilde{X}}
\newcommand{\tY}{\tilde{Y}}

\newenvironment{prooflem}[1]{\noindent \textbf{Proof of Lemma \ref{#1}:}}{\hfill $\Box$ \\}

\newenvironment{proofprop}[1]{\noindent \textbf{Proof of Proposition
\ref{#1}:}}{\hfill $\Box$ \\}

\newcommand{\Sigmahat}{\hat{\Sigma}}
\newcommand{\Gammahat}{\hat{\Gamma}}
\newcommand{\Phihat}{\hat{\Phi}}

\newcommand{\Pbf}{{\mathbf P}}

\newcommand{\M}{{\mathbf M}}

\newcommand{\xit}{\tilde{\xi}}
\newcommand{\etat}{\tilde{\eta}}

\newcommand{\sigmat}{\tilde{\sigma}}

\newcommand{\yspc}{\mathcal Y}
\newcommand{\xspc}{\mathcal X}
\newcommand{\rspc}{\mathcal R}

\newcommand{\bspc}{{\mathcal B}}

\newcommand{\espc}{{\mathcal E}}

\newcommand{\noboxfig}[2]{\includegraphics[width=#2in]{#1.eps}}
\newcommand{\boxfig}[2]{\fbox{\noboxfig{#1}{#2}}}

\newcommand{\smallfig}{2.38}

\newcommand{\Dfig}[2]{\noboxfig{#1}{#2}}
\newcommand{\Cfig}[2]{\centerline{\psfig{figure=#1.eps,width=#2in}}}

\begin{document}

\begin{frontmatter}

\title{Fundamentals of  path analysis in the social sciences}

 \author[stat]{R. Dennis Cook}%

\author[fiq]{Liliana Forzani \corref{cor1}}%
 
\cortext[cor1]{Corresponding author. Facultad de Ingenier\'ia Qu\'imica, UNL. Researcher
of CONICET, Santa Fe, Argentina. Email: liliana.forzani@gmail.com.}

\address[stat]{School of Statistics, University
of Minnesota, Minneapolis, MN 55455. E-mail: dennis@stat.umn.edu}
\address[fiq]{Corresponding author. Facultad de Ingenier\'ia Qu\'imica, UNL. Researcher
of CONICET, Santa Fe, Argentina. Email: liliana.forzani@gmail.com}

\begin{abstract}
 Motivated by a recent series of diametrically opposed articles on the relative value of statistical methods for the analysis of path diagrams in the social sciences, we discuss from a primarily theoretical perspective selected fundamental aspects of  path modeling and analysis based on a common reflexive setting.  Since there is a paucity of technical support evident in the debate, our aim is to connect it to mainline statistics literature and to address selected foundational issues that may help move the discourse.  We do not intend to advocate for or against a particular 
 method or analysis philosophy.
\end{abstract}

%


\end{frontmatter}


\section{Introduction}\label{sec:intro}

Path modeling has become a standard for representing social science theories graphically.  The validation of a path model often involves inference about concepts like ``customer satisfaction'' or ``competitiveness'' for which there are no objective measurement scales.  Since such concepts cannot be measured directly, multiple surrogates, which are often called indicators, are  used to gain information about them indirectly.  The role of a path diagram is then to provide a visual representation of the relationships between the concepts represented by latent variables and the indicators.  We found the discussion by \citet{Rigdon2012} to be  helpful for gaining an overview of this general paradigm.

There are two analysis protocols, referred to broadly as partial least squares (PLS) path analysis and path analysis by structural equation modeling (SEM), that have for decades been used to estimate and infer about unknown quantities in a path diagram.  The advocates for these two protocols now seem at loggerheads over the relative advantages and disadvantages of the methods, due in part to  the publication of two recent articles \citep*{Ronkko2013, Ronkko2016} that are highly critical of PLS methods for path analysis.
Within the context of path modeling, \citet{Ronkko2016}
asserted that ``\ldots the majority of claims made in PLS literature should be categorized as statistical and methodological myths and urban legends,'' 
and recommended that the use of ``\ldots PLS should be discontinued until the methodological problems explained in [their] article have been fully addressed.''   In response, proponents of PLS appealed for a more equitable  assessment.  Henseler and his nine coauthors  \citep{Henseler2014} tried to reestablish a constructive discussion on the role PLS in path analysis  while arguing that, rather than moving the debate forward, \citet{Ronkko2013} created new myths.  \citet*{Mcintosh2014} attempted to  resolve the issues separating \citet{Ronkko2013} and \citet{Henseler2014} in an `even-handed manner,' concluding by lending their support to Rigdon's  \citep{Rigdon2012} recommendation that steps be taken to divorce PLS path analysis from its SEM competitors by developing methodology from a purely PLS perspective.  The appeal for balance was evidently found wanting by the  Editors of {\em The Journal of Operations Management}, who established a policy of desk-rejecting papers that use PLS \citep{Guide2015}, and by \citet{Ronkko2016} who restated and reinforced the views of \citet{Ronkko2013}.  Asserting that part of the debate rests with differential appreciations of concepts, \citet*{Sarstedt2016} offered a unifying framework intended to disentangle the confusion between the terminologies and  thereby facilitate resolution or at least a common understandings.  They also argue that PLS is preferable in common settings.  \citet*{Akter2017} concluded that PLS path modeling is a promising tool for dealing with complex models encountered in big data analytics \citep[see also][]{CookForzani2017PLS, CookForzani2018}.

There is a substantial literature that bears on this debate and we have not attempted a comprehensive review (\citet{Ronkko2016} cites over 150 references).  Nevertheless, our impression is that nearly all articles are written from a practitioners standpoint,  relying mostly on intuition and simulations to support sweeping statements about intrinsically mathematical/statistical issues without adequate supporting theory.   Not being steeped in the PLS path modeling culture, we found  the literature to be a bit challenging because of the implicit expectation that the reader would understand the implications of various customs and terminology.  Few give a careful statement of the ultimate statistical objectives of SEM or PLS path modeling, or define terms like ``bias'', which seems to be used by some in a manner inconsistent with its widely-accepted meaning in statistics.  And disagreements have arisen even when care is taken.  For instance, \citet{Mcintosh2014} took issue with Henseler, et al.'s (2014) description of a ``composite factor model'' arguing that it is instead a ``common factor model.'' 
 It seems to us that adequately addressing the methodological issue in path analysis requires avoiding such ambiguity  by employing a degree of context-specific theoretical specificity.  
In this article we appeal to the theoretical foundations of PLS established by \citet*{CookHellandSu2013} to hopefully shed light on some of the fundamental issues in the PLS v. SEM debate.    Our intention is not to advocate for or against either methodology, but we hope that our discussion will help the debate. Our attempt at balance seems appropriate because some \citep[eg.][]{Sarstedt2016, Akter2017} evidently do not share the hard  views represented by \citet{Ronkko2013} and \cite{Ronkko2016}.

It may be helpful to recognize that the foundations of PLS path analysis are intertwined with those of PLS regression, as it will be relevant later in this article.

\subsection{PLS regression}
PLS regression has a long history going back at least to \citet*{Wold1983}, although PLS-type methods  have been traced back to the mid 1960s \citep{Geladi1988}.  It is one of the first methods for prediction in high-dimensional multi-response linear regressions  in which the sample size $n$ may  not be large relative to the number of predictors $p$ or the number of responses $r$. Although studies have appeared in statistics literature from time to time \citep[eg.][]{Helland1990,Helland1992,Frank1993,Chun2010,CookHellandSu2013}, the development of PLS regression has taken place mainly within the Chemometrics community where empirical prediction is a central issue and PLS regression is now a core method.    PLS regression now has a substantial following across the applied sciences. The two main aspects of PLS regression that have made it appealing are its ability to give useful predictions in high-dimensional regressions and to provide valuable insights through graphical interpretation of the  loadings that yield the PLS composites.  However, its algorithmic nature evidently made it inconvenient to study using traditional statistical measures, with the consequence that PLS regression was long regarded as a technique that is useful, but whose core statistical properties and operating characteristics are elusive.  This ambiguity may have contributed to the recommendation by \citet[][end of Section 1]{Ronkko2016} that PLS path modeling should be suspended until its theoretical and methodological problems are sorted out.  

 Standard statistical properties of PLS regressions were largely unknown until the recent work of \cite{CookHellandSu2013}.  They showed that the most common PLS regression algorithm, SIMPLS \citep{deJong1993}, produces a moment-based $\sqrt{n}$-consistent estimator of a certain population subspace.  Called an envelope, that population subspace is designed to envelop the material information in the data and thereby exclude the immaterial information, resulting in estimators with less variation asymptotically.  In multi-response linear regression envelopes can be applied to the responses, the predictors or both simultaneously.  \cite{CookHellandSu2013} also produced the first statistically firm model for PLS regression and showed that likelihood-based estimators outperform traditional moment-based PLS regression estimators even when  assumptions leading to the likelihood are violated.  \citet{CookForzani2017PLS, CookForzani2018} showed that in high-dimensional regressions PLS predictions can converge at the $\sqrt{n}$ rate regardless of the relationship between $n$ and $p$.  

\subsection{Outline}

 To establish a degree of context-specific theoretical specificity, we cast our development in the framework of a common reflexive path model that is stated in Section~\ref{setup}.  This model covers simulation results by \citet[][Figure 1]{Ronkko2016}, which is one reason for its adoption.
We show in Section~\ref{sec:RRR} that the apparently novel observation that this setting implies a reduced rank regression (RRR) model   for the observed variables  \citep*{Anderson1951, Izenman:1975aa, ransey, Cook:2015ab}. From here we address identifiability and estimation using RRR. The approach we take and the broad conclusions that we reach should be applicable to other perhaps more complicated contexts.  Some of the concerns expressed by \citet{Ronkko2016} involve contrasts with SEM methodology.  Building on RRR, we introduce SEM methodology into the discussion in Section~\ref{SEM} and show that under certain key assumptions additional parameters are identifiable under the SEM  model.
 In Section~\ref{bias} we quantify the bias and revisit the notion of consistency at large.  
   
Having established the context and background, we turn to  PLS path modeling in Section~\ref{sec:PLS}, describing how envelopes  \citep*{Cook2010} play a central role in characterizing its behavior.  This will then allow us to address some of the main issues in the PLS v. SEM debate.  A few simulation results are presented in Section~\ref{sec:example}, and some myths and urban legends described by \citet{Ronkko2016} are addressed in Section~\ref{sec:myths}, including (a) the role of PLS in reducing estimative variation, (b) the meaning of the PLS weights and their optimality, (c) bias, (d) usefulness of PLS in small samples, (e) how PLS has the potential to reduce measurement error through its treatment of the variance/covariance structure, and (f) versions of PLS estimators based on likelihood theory.  Many technical details have been placed in appendices.

\section{A reflexive path model}\label{setup}

\begin{figure}[ht]
\begin{center}
\includegraphics[width=0.68\textwidth]{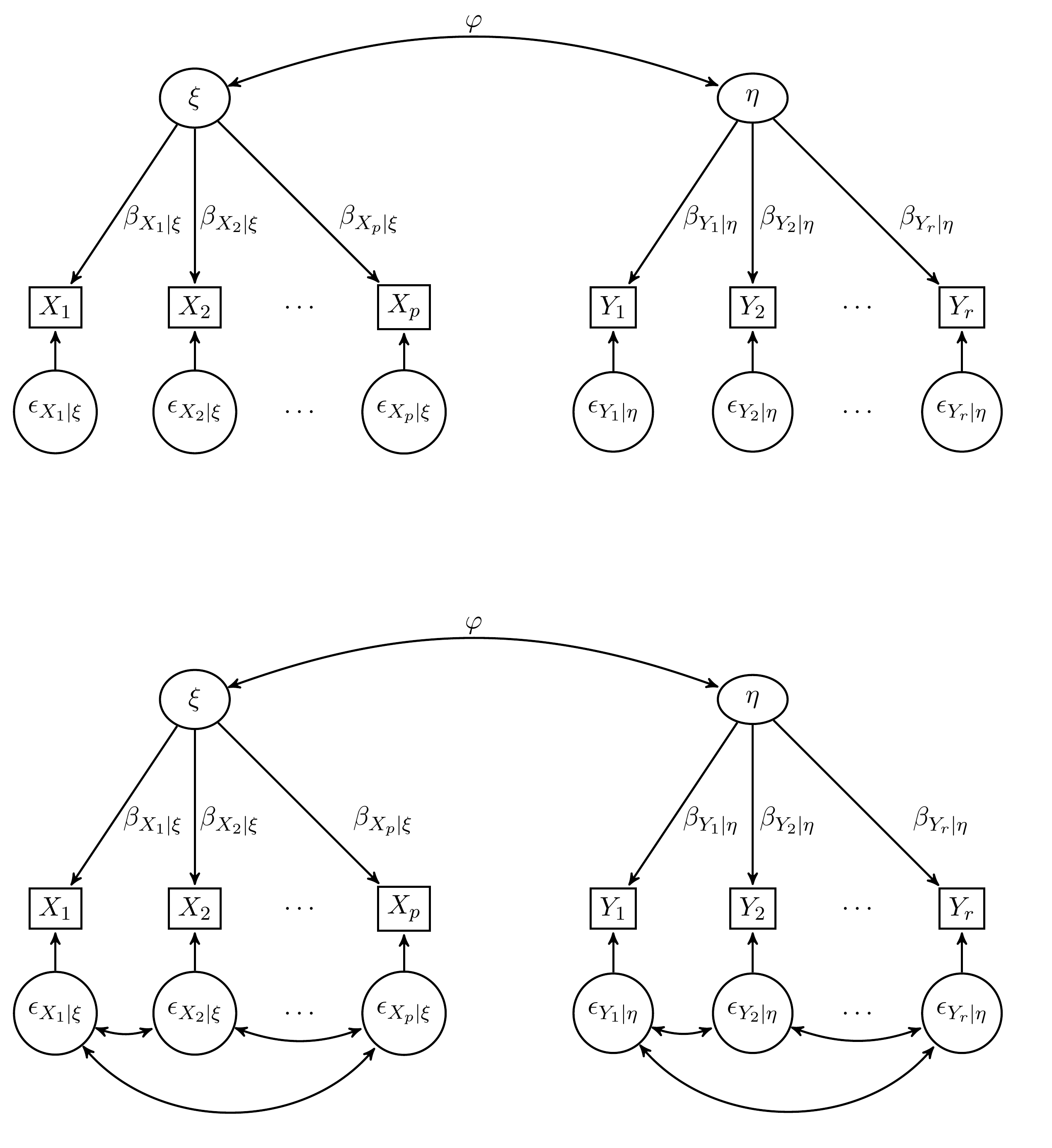}  
\caption{Reflexive path diagrams relating two latent constructs $\xi$ and $\eta$ with their respective univariate indicators, $X_{1},\ldots,X_{p}$ and $Y_{1},\ldots,Y_{r}$.  $\varphi$ denotes either $\cor(\xi, \eta)$ or 
$\cor\{E(\xi \mid X), E(\eta \mid Y)\}$.}  \label{figpath}
\end{center}
\end{figure}

 The upper and lower path diagrams presented in Figure~\ref{figpath}  are  typical  reflexive path models \citep[eg.][]{LohmllerJ.-B.1989}. 
In each diagram elements of $Y=(Y_{j}) \in \real{r}$, $X = (X_{j})\in \real{p}$ denote observable random variables, which are often called indicators, that are assumed to provide reflexive information about underlying latent constructs $\xi, \eta \in \real{}$.  This is indicated by the arrows in Figure~\ref{figpath} leading from the constructs to the indicators.  Boxed variables are observable while circled variables are not.   This restriction to univariate constructs is common practice in  the social sciences.  An important path modeling goal in this setting is to infer about the association between the latents $\eta$ and $\xi$, as indicated by the double-headed curved paths between the constructs. We take this to be a general objective of both PLS and SEM, and it is one focal point of  this article.  Each indicator is affected by an unobservable error $\epsilon$.  The absence of paths between these errors indicates that, in the upper path diagram, they are independent conditional on the latent variables.  They are allowed to be conditionally dependent in the lower path diagram, as indicated by the double-arrowed paths that join them.  The coefficients (not shown) for these paths represent the covariances between the errors.

For instance, \citet[][p. 228--235, Fig. 7.3]{Bollen1989} described a case study on the evolution of political democracy in developing countries.  The variables $X_{1},\ldots,X_{p}$ were indicators of specific aspects of a political democracy in developing countries in 1960, like measures of the freedom of the press and the fairness of elections.  The latent construct $\xi$ was seen as a single combined latent construct representing the level of democracy 1960.  The variables $Y_{1},\ldots,Y_{r}$ were the same indicator from the same source  in 1965, and correspondingly $\eta$ was interpreted as a single latent construct that measures the level of democracy in 1965.

Also, the path diagrams in Figure~\ref{figpath} are uncomplicated relative to those that may be used in sociological studies. For example, \citet*{Vinzi2010} described a study of fashion in which an instance of the upper diagram of Figure~\ref{figpath} was imbedded in a larger path diagram  with the latent constructs ``Image'' and ``Character'' serving as reflexive indicators of a third latent construct ``Brand Preference''  with its own indicators.  Our discussion centers largely on the path diagrams of Figure~\ref{figpath} since these seem sufficient to allow informative comparisons of PLS and SEM.

Assuming that all effects are additive, the upper and lower path diagrams in Figure~\ref{figpath} can be represented as a common multivariate (multi-response) regression model:
\begin{eqnarray}
\left( \begin{array}{c}\xi \\ \eta \end{array} \right) &=&
\left( \begin{array}{c}\mu_{\xi} \\ \mu_{\eta} \end{array} \right)
 + \epsilon_{\xi,\eta},\mathrm{\;where\;}  \epsilon_{\xi,\eta} \sim N_{2}(0,\Sigma_{\xi,\eta}). \nonumber \label{model-1}\\
 \vspace{.02in}\nonumber \\
  Y &=& \mu_{Y} + \beta_{Y|\eta} (\eta -\mu_{\eta})+ \epsilon_{Y|\eta} ,  \mathrm{\;where\;} \epsilon_{Y|\eta} \sim N_{r}(0,\Sigma_{Y| \eta}).\label{model-2} \\
 X & = & \mu_{X } +\beta_{X|\xi} (\xi-\mu_{\xi}) + \epsilon_{X|\xi} ,  \mathrm{\;where\;} \epsilon_{X|\xi} \sim N_{p}(0, \Sigma_{X| \xi}). \label{model-3}
\end{eqnarray}
We use $ \sigma_{\xi}^{2}$, $\sigma_{\xi \eta}$ and $\sigma^2_{\eta}$ to denote the elements of $\Sigma_{\xi,\eta} \in \real{2 \times 2}$, and we further assume that $ \epsilon_{\xi,\eta}$, $ \epsilon_{X|\xi}$ and $ \epsilon_{Y|\eta}$ are mutually independent so jointly $X$, $Y$, $\xi$ and $\eta$ follow a multivariate normal distribution. We occasionally address this normality assumption but do not dwell on it because it seems to be a secondary consideration in view of other overarching structural issues.  We use the notation $\Sigma_{A B}$ to denote the matrix of covariances between the elements of the random vectors $U$ and $V$, and we use $\beta_{U|V} = \Sigma_{U V}\Sigma_{V}^{-1}$ to indicate the matrix of population coefficients from multi-response regression of $U$ on $V$, where $\Sigma_{V} = \var(V)$.  Appendix  Lemma~\ref{lemma-one} gives the joint multivariate mean and variance of $X$, $Y$, $\xi$ and $\eta$.  

 \citet{Henseler2014}  refers to this as a {\em common factor model} when the $\Sigma_{X|\xi}$ and $\Sigma_{Y|\eta}$ are taken to be diagonal matrices, corresponding to the upper diagram in Figure~\ref{figpath}, and as a {\em composite factor model} when $\Sigma_{X|\xi}$ and $\Sigma_{Y|\eta}$ are not constrained to be diagonal,  corresponding to the lower diagram in Figure~\ref{figpath}, although these designations may not be universal.  \citet{Mcintosh2014} refers to both as common factor models with accompanying descriptions of $\Sigma_{X|\xi}$ and $\Sigma_{Y|\eta}$. 
In contrast to the composite factor model described by \citet{Henseler2014}, the composite factor model described by  \citet[][p. 216]{Mcintosh2014} takes the constructs to be exact linear combinations of the indicators, $\eta = \beta_{\eta}^{T}Y$, $\xi = \beta_{\xi}^{T}X$.  The model then requires that the indicators be independent given $\eta$ and $\xi$ -- $Y \indep X \mid \beta_{\eta}^{T}Y, \beta_{\xi}^{T}X$ -- so the constructs are responsible for any dependence between $X$ and $Y$.  The definition of the common factor model by \citet{Henseler2014} seems to be the most widely accepted, so we confine attention to it until Section~\ref{sec:myths} where we offer introductory contrasts with the model by \citet{Mcintosh2014}.

\citet{Rigdon2012} questioned whether constructs like $\xi$ and $\eta$ are necessarily the same as the theoretical concepts they are intended to capture or whether they are approximations of the theoretical concepts allowed by the selected indicators.  It the latter view is taken, then an understanding of the constructs cannot be divorced from the specific indicators selected for the study.  This view will be relevant later in this section when considering ways of measuring association between $\xi$ and $\eta$.

 The constructs $\eta$ and $\xi$ are not well-defined since any non-degenerate linear transformations of them lead to an equivalent model and, in consequence, it is useful to introduce harmless constraints to facilitate estimation, inference and interpretation.  To this end, we consider two sets of constraints:
  \begin{eqnarray*}
 \mathrm{Regression\;  constraints:} & & \mu_{\xi} = \mu_{\eta} = 0 \mathrm{\;and \;} \var\{\E(\xi \mid X)\} = \var\{\E(\eta \mid Y)\} = 1.\\
 \mathrm{Marginal \; constraints:}  & & \mu_{\xi} = \mu_{\eta} = 0 \mathrm{\;and \;} \var(\xi) = \var(\eta) = 1.
 \end{eqnarray*}
 The regression constraints fix the variances of the conditional means $ \E(\xi \mid X)$ and $\E(\eta \mid Y)$ at 1, while the marginal constraints fix the marginal variances of $\xi$ and $\eta$ at 1.  The regression and marginal constraints are related via the variance decompositions  $\var(\xi) =   \var\{\E(\xi \mid X)\} + \E\{\var(\xi \mid X)\}$ and $\var(\eta) =   \var\{\E(\eta \mid Y)\} + \E\{\var(\eta \mid Y)\}$.  It seems to us that the choice of constraints is mostly a matter of taste: We will show later that for the SEM model that choice has no impact on the end result.

 The association between $\eta$ and $\xi$ has been measured in the path modeling literature via the correlation $\cor \{\E (\eta \mid Y), \E(\xi \mid X)\}$, which is  sometimes called an ideal weight  \citep{Ronkko2016}, or via the correlation $\cor (\eta, \xi)$. We reason that these measures can reflect quite different views of a reflexive setting.  The marginal correlation $\cor (\eta, \xi)$ implies to us that $\eta$ and $\xi$ represent concepts  that are uniquely identified regardless of any subjective choices made regarding the variables $Y$ and $X$ that are reasoned to reflect their properties up to linear transformations. Two investigators studying the same constructs would naturally be estimating the same correlation even if they selected a different $Y$ or a different $X$.  In contrast,  $\cor \{\E (\eta\mid Y), \E(\xi\mid X)\}$, the correlation between population regressions, suggests a conditional view:  $\eta$ and $\xi$ exist only by virtue of the variables that are selected to reflect their properties and so attributes of $\eta$ and $\xi$ cannot be claimed without citing the corresponding $(Y, X)$.   Two investigators studying the same concepts could understandably be estimating different correlations if they selected a different $Y$ or a different $X$: 
 $\cor 
 \{ \E (\eta\mid Y), \E(\xi\mid X)\} \neq \cor \{ \E (\eta^{*}\mid Y^{*}), \E(\xi^{*}\mid X^{*}) \}$.  For instance, the latent construct ``happiness'' might reflect in part a combination of happiness at home and happiness at work.  Two sets of indicators $X$ and $X^{*}$ that reflect these happiness subtypes differently might yield different correlations with a second latent construct, say ``generosity.''   The use of $\cor \{\E (\eta \mid Y), \E(\xi \mid X)\}$ as a measure of association can be motivated also by an appeal to dimension reduction.  Under the models (\ref{model-2}) and (\ref{model-3}), $X \indep \xi \mid \E(\xi \mid X)$ and $Y \indep \eta \mid \E(\eta \mid Y)$.  In consequence, the constructs affect their respective indicators only through the linear combinations given by the conditional means, $\E(\xi \mid X) = \Sigma_{X\xi}^{T}\Sigma_{X}^{-1}(X - \mu_{X})$ and $\E(\eta \mid Y) = \Sigma_{Y\eta}^{T}\Sigma_{Y}^{-1}(Y-\mu_{Y})$.

\section{Reduced rank regression}\label{sec:RRR}
PLS path algorithms gain information on the latent constructs from the multi-response linear regressions of $Y$ on $X$ and $X$ on $Y$.  Consequently, it is useful to provide a characterization of these regressions.  In the following lemma we state that for the model presented in Section 2, the regression of $Y$ on $X$ qualifies as a reduced rank regression (RRR). The essential implication of the lemma is that, under the reflexive model described in Section~\ref{setup}, the coefficient matrix $\beta_{Y|X}$ in the multi-response linear regression of $Y$ on $X$ must have rank 1 because $\cov(Y,X) = \Sigma_{YX}$ has rank 1.  A similar conclusion applies to the regression of $X$ on $Y$.

\begin{lemma} \label{RRR}
For model presented in Section \ref{setup} and without imposing either the regression or marginal constraints, we have
$
E (Y \mid X) 
= \mu_Y  + \beta_{Y|X} (X - \mu_X), 
$
where $\beta_{Y|X} = A B^T \Sigma_{X}^{-1}$, and $A \in {\mathbb R}^{r \times 1} $ and $B\in {\mathbb R}^{p \times 1}$ are such that $\Sigma_{YX}  = A B^T  $.
\end{lemma}

It is known from the literature on RRR that the vectors $A$ and $B$ are not identifiable while $A B^T$ is identifiable.
As a consequence of being a reduced rank regression model, we are  able to state in Proposition \ref{one} which parameters are identifiable  in the reflexive model presented in Section \ref{setup}. 

\begin{prop}\label{one}
The parameters $\Sigma_X$, $\Sigma_Y$, $\Sigma_{YX}=A B^T $, $\beta_{Y|X}$, $\mu_Y $ and $\mu_X$ are identifiable in the reduced rank model of Lemma~\ref{RRR}, but $A$ and $B$ are not. 
Additionally, under the regression constraints, 
 the quantities
$ \sigma_{\xi \eta}/(\sigma^2_{\eta} \sigma^2_{\xi})$,   $cor \{E( \xi|X),E(\eta|Y)\}$, $E( \xi|X) $, $E(\eta|Y)$, $\Sigma_{X \xi}$ and $\Sigma_{Y \eta }$ are identifiable in the reflexive model except for  sign, while $\cor(\eta, \xi)$,
$\sigma^2_{\xi}$, $\sigma^2_{\eta}$, $\sigma_{\xi \eta}$,  $\beta_{X|\xi}$, $ \beta_{Y|\eta}$,
$\Sigma_{Y|\eta} $ and $\Sigma_{X|\xi}$ are not identifiable.  Moreover,
\begin{align}
|\cor \{E( \xi|X),E(\eta|Y)\}| 
& =   |\sigma_{\xi \eta}|/(\sigma^2_{\eta} \sigma^2_{\xi}) = \tr^{1/2}(\beta_{X|Y}\beta_{Y|X}) = \tr^{1/2}(\Sigma_{XY}\Sigma_{Y}^{-1}\Sigma_{YX}\Sigma_{X}^{-1}) \label{cor}  
\end{align}
where  $|\cdot|$ denotes the absolute value of its argument.
 \end{prop}
 
Although $A$ and $B$ are not identifiable, $|\cor \{E( \xi|X),E(\eta|Y)\}|$ is identifiable since from (\ref{cor}) it depends on the identifiable quantities $\Sigma_{X}$, $\Sigma_{Y}$ and on the rank 1 covariance matrix $\Sigma_{YX} = AB^{T}$ in the reduced-rank regression of $Y$ on $X$.   Let $\Sigmahat_{YX}$ denote the maximum likelihood estimator of $\Sigma_{YX}$ from fitting the reduced rank regression model of Lemma~\ref{RRR}, and let $\Sigmahat_{X}$ and $\Sigmahat_{Y}$ denote the sample versions of $\Sigma_{X}$ and $\Sigma_{Y}$ \citep{Cook:2015ab}.  Then the maximum likelihood estimator of $|\cor \{E( \xi|X),E(\eta|Y)\}|$ can be obtained by substituting these estimators into (\ref{cor}):
  \[
 |\widehat{\cor} \{E( \xi|X),E(\eta|Y)\}| = \tr^{1/2}(\Sigmahat_{XY}\Sigmahat_{Y}^{-1}\Sigmahat_{YX}\Sigmahat_{X}^{-1}).
 \]

Let $\Sigma_{\tY\tX} = \Sigma_{Y}^{-1/2}\Sigma_{YX}\Sigma_{X}^{-1/2}$ denote the standardized version of $\Sigma_{Y X}$ that corresponds to the rank one regression of $\tilde{Y}=\Sigma_{Y}^{-1/2} Y$ on $\tilde{X}=\Sigma_{X}^{-1/2}X$.  Then $|\cor \{E( \xi|X),E(\eta|Y)\}|$ can be written also as
\begin{equation}\label{est}
|\cor \{E( \xi|X),E(\eta|Y)\}| = \tr^{1/2}(\Sigma_{\tX\tY}\Sigma_{\tY\tX}) = \|\Sigma_{\tY\tX}\|_{F},
\end{equation}
which can be seen as the Frobenius norm $\|\cdot\|_{F}$ of the standardized covariance matrix $\Sigma_{\tY\tX}$.

The maximum likelihood estimator  $|\widehat{\cor} \{E( \xi|X),E(\eta|Y)\}| $ can be computed in the following steps starting with the data $(Y_{i},X_{i})$, $i=1,\ldots,n$ \citep{Cook:2015ab}:
\begin{enumerate}
\item Standardize  $\tilde{Y}_{i} = \Sigmahat_{Y}^{-1/2}Y_{i}$ and $\tilde{X}_{i} = \Sigmahat_{X}^{-1/2}X_{i}$, $i=1,\ldots,n$.
\item Construct $\Sigmahat_{\tY\tX}$, the matrix of sample correlations between the elements of the standardized vectors $\tX$ and $\tY$. 
\item Form the singular value decomposition $\Sigmahat_{\tY \tX} = U D V^{T}$ and extract
$
\Sigmahat_{\tY \tX}^{(1)} = U_{1}D_{1}V_{1}^{T},
$
where $U_{1}$ and $V_{1}$ are  the first columns of $U$ and $V$, and $D_{1}$ is the corresponding (largest) singular value of $\Sigmahat_{\tY \tX}$.  

\item Then 
\[
 |\widehat{\cor} (E( \xi|X),E(\eta|Y))| = \|\Sigmahat_{\tY \tX}^{(1)}\|_{F} =  D_{1}.
 \]
 In short, $ |\widehat{\cor} \{E( \xi|X),E(\eta|Y)\}|$ is the largest singular value of $\Sigmahat_{\tY \tX}$.  Equivalently it is the first sample canonical correlation between $X$ and $Y$.
\end{enumerate}

\section{SEM}\label{SEM}
As was stated in Proposition \ref{one}, the quantity $\cor (\eta, \xi)$ is not identifiable without further assumptions. We give now sufficient conditions to have identification of $\cor (\xi, \eta)$. The conditions needed are related to the identification of $\Sigma_{X|\xi}$ and $\Sigma_{Y|\eta}$.  As seen in Appendix Lemma~\ref{aux},
\begin{eqnarray} 
\Sigma_X &=& \Sigma_{X|\xi} + \sigma_{\xi}^{-2} (B^T\Sigma_X^{-1} B)^{-1}  B B^T \label{a}\\
\Sigma_Y &=& \Sigma_{Y|\eta} + \sigma_{\eta}^{-2} (A^T\Sigma_Y^{-1} A)^{-1}  A A^T. \label{c}
 \end{eqnarray}
 The terms $(B^T\Sigma_X^{-1} B)^{-1}BB^{T}$  and $ (A^T\Sigma_Y^{-1} A)^{-1}  A A^T$ in (\ref{a}) and (\ref{c}) 
 are identifiable.  Since the goal of SEM is to estimate $\cor(\xi, \eta)$ and since we know from Proposition~\ref{one} that 
 \begin{equation}\label{corE}
| \cor\{E(\xi \mid X), E(\eta \mid Y)\} |=  |\sigma_{\xi \eta}|/(\sigma_{\xi}^{2}\sigma_{\eta}^{2})
 \end{equation}
  is identifiable, we will be able to identify $|\cor(\xi,\eta)|$ if we can identify $\sigma_{\xi}^{2}$ and $\sigma_{\eta}^{2}$.  From (\ref{a}) and (\ref{c}) that is equivalent to identifying $\Sigma_{Y|\eta}$ and $\Sigma_{X|\eta}$.  We show in Appendix Proposition~\ref{two} that (a) if $\Sigma_{Y|\eta} $ and $\Sigma_{X|\xi}$ are identifiable then $\sigma_{\xi}^{2}$, $\sigma_{\eta}^{2}$, $|\sigma_{\xi \eta}|$,  $\beta_{X|\xi}$, $ \beta_{Y|\eta}$ are identifiable and that (b) $\Sigma_{Y|\eta} $ and $\Sigma_{X|\xi}$ are identifiable if and only if $\sigma_{\xi}^{2}$, $\sigma_{\eta}^{2}$ are so.  

%
%

The next proposition gives conditions that are sufficient to ensure identifiability.  Let $(M)_{ij}$ denote the $ij$-th element of the matrix $M$ and $(V)_{i}$ the $i$-th element of the vector $V$.
\begin{prop}\label{prop:three1}
Under the regression constraints,
(I) if $ \Sigma_{X|\xi}$ contains an off-diagonal element that is known to be zero, say $( \Sigma_{X|\xi})_{ij} =0$, and if $(B)_{i}(B)_{j} \neq 0$ then $\Sigma_{X|\xi}$ and $\sigma_{\xi}^{2}$ are identifiable.
(II) if $ \Sigma_{Y|\eta}$ contains an off-diagonal element that is known to be zero, say $( \Sigma_{Y|\eta})_{ij} = 0$, and if $(A)_{i}(A)_{j} \neq 0$ then $\Sigma_{Y|\eta}$ and $\sigma_{\eta}^{2}$ are identifiable.

\end{prop}

\begin{Cor}\label{cor:three1}
Under the regression constraints,
if $\Sigma_{Y|\eta}$ and $ \Sigma_{X|\xi}$ are diagonal matrices and if $A$ and $B$ each contain at least two non-zero elements then   $\Sigma_{Y|\eta} $, $\Sigma_{X|\xi}$, $\sigma_{\xi}^{2}$ and $\sigma_{\eta}^{2}$ are identifiable.
\end{Cor}

The usual assumption in the SEM model is that $\Sigma_{X|\xi}$ and $\Sigma_{Y|\eta}$ are diagonal matrices \citep[eg.][]{joreskog1970,Henseler2014}.  We see from (\ref{corE}) and Corollary~\ref{cor:three1} that this assumption along with the regression constraints is sufficient to guarantee that $|\cor(\xi, \eta)|$ is identifiable provided $B$ and $A$ contain at least two non-zero elements.  However, from Proposition~\ref{prop:three1} we also see that is not necessary for $\Sigma_{Y|\eta}$ and $ \Sigma_{X|\xi}$ to be diagonal.  The assumption that $\Sigma_{X|\xi}$ and $\Sigma_{Y|\eta}$ are diagonal matrices means that, given $\xi$ and $\eta$, the elements of $Y$ and $X$ must be independent.  In consequence, elements of $X$ and $Y$ are correlated only by virtue of their association with $\eta$ and $\xi$.   The presence of any residual correlations after accounting for $\xi$ and $\eta$ would negate the model and possibly lead to spurious conclusions.  See \citet{Henseler2014} for a related discussion.

In full, the usual SEM model requires that $\Sigma_{Y|\eta} = D_{Y|\eta}$ and $ \Sigma_{X|\xi}=D_{X|\xi}$ are diagonal matrices, $D_{Y|\eta}$ and $ D_{X|\xi}$, and it adopts the marginal constraints instead of the regression constraints.  
 By Proposition \ref{prop:three1}, our ability to identify parameters is unaffected by the constraints adopted.
   However, we need also to be sure that the meaning of $\cor (\xi, \eta) $ is unaffected by the constraints adopted.

\begin{lemma}\label{sem-conditions}
In the SEM model, $\cor (\xi, \eta)$ is unaffected by choice of constraint,  $\sigma_{\xi}^{2}= \sigma_{\eta}^{2} =1$ or $\var\{E(\xi|X)\} = \var\{E(\eta|Y)\} =1$.
\end{lemma}

Now, to estimate  the identifiable parameters with the regression or marginal constraints, we take the joint distribution of $(X,Y)$ and  insert the conditions imposed. The maximum likelihood estimators can then be found under normality, as shown in Appendix Proposition~\ref{computer}.  
From this we conclude that the same maximization problem arises under either sets of constraints.

\section{Bias}\label{bias}
We understand that the notion of bias in the literature on path analysis refers to the difference between measuring the association via $ \cor \{E( \xi|X),E(\eta|Y)\}$ with the regression constraints and via $\cor(\xi, \eta)$ with the marginal constraints.  To establish this connection, we let $\xi$ and $\eta$ denote the latent variables under the marginal constraints, so $\var(\xi) = \var(\eta) = 1$, and let $\xit = \xi\var^{-1/2}\{E(\xi \mid X)\} $ and 
$\etat = \eta \var^{-1/2}\{E(\eta \mid Y)\} $ denote the corresponding latent variables with the regression constraints, so $\var\{E(\xit \mid X)\} = \var\{E(\etat \mid Y)\} = 1$.  Then (See Appendix~\ref{sec:bias})
\begin{eqnarray*}
\cor\{E(\xit \mid X), E(\etat \mid Y)\} 
& = & [\var\{E(\xi \mid X)\}  \var\{E(\eta \mid Y)\} ]^{1/2}\cor(\xi, \eta).
\end{eqnarray*}
To interpret this results we use the decomposition
\[
\var(\xi) = 1 = E\{\var(\xi \mid X)\} + \var\{E(\xi \mid X)\} = \var(\xi \mid X) + \var\{E(\xi \mid X)\},
\]
where the final equality holds because under normality $\var(\xi \mid X)$ is non-stochastic.  It follows from this representation that $ \var\{E(\xi \mid X)\} \leq1$ and $ \var\{E(\eta \mid Y)\} \leq 1$.   In consequence, $|\cor\{E(\xit \mid X), E(\etat \mid Y)\}|$ is a lower bound on  $|\cor(\xi, \eta)|$:
\begin{equation}\label{corrbnd}
|\cor\{E(\xit \mid X), E(\etat \mid Y) \}| \leq  |\cor(\xi, \eta)|
\end{equation} 

The term $\var(\xi \mid X) $ represents how well $X$ predicts $\xi$.   If this term is small relative to $\var(\xi) = 1$, then $ \var\{E(\xi \mid X)\}$ will be close to $1$, and in consequence 
\[
|\cor\{E(\xit \mid X), E(\etat \mid Y)\}| \approx [  \var\{E(\eta \mid Y)\} ]^{1/2}|\cor(\xi, \eta)|.
\]
The conditional variance $\var(\xi \mid X) $ will be small when the signal is strong.  This will happen with few highly informative predictors.  It will also happen as number of informative indicators  increases, a scenario that is referred to as an {\em abundant} regression in statistics \citep{Cook2012estimating,Cook2013}. Repeating this argument for $\var(\eta \mid Y)$ we see that 
\[
|\cor\{E(\xit \mid X), E(\etat \mid Y)\}| \approx |\cor(\xi, \eta)|
\]
when $X$ and $Y$ are such that $\var(\xi \mid X)$ and $\var(\eta \mid Y)$ are both small relative to 1, which is often referred to as  consistency-at-large in the literature on path analysis \citep{Hui1982}.  Sufficient conditions for this to occur are that (a) the eigenvalues of $\Sigma_{X|\xi}$ and $\Sigma_{Y|\eta}$ are bounded away from $0$ and $\infty$ as $p, r \rightarrow \infty$ and (b) $\Sigma_{X \xi}^{T}\Sigma_{X \xi}$ and $\Sigma_{Y \eta }^{T}\Sigma_{Y \eta}$ both diverge as $p, r \rightarrow \infty$.  This result seems to align well with the argument given at the end of Section~\ref{setup}:  $\cor\{E(\xit \mid X), E(\etat \mid Y)\}$ and $\cor(\xi, \eta)$ agree when $X$ and $Y$ provide (nearly) exhaustive information about $\xi$ and $\eta$.  The crucial SEM assumption that  $\Sigma_{X|\xi}$ and $\Sigma_{Y|\eta}$ are diagonal matrices plays no role in this conclusion.

\section{PLS }\label{sec:PLS}

Throughout most if its history, PLS meant empirical algorithms that lacked an underlying Fisherian statistical model that could serve to characterize its performance and goals in terms of population parameters. However, it is now known that PLS algorithms are moment-based methods that estimate a population subspace called an envelope that is associated with a particular class of regression models.  This recent advance means that PLS can now be studied statistically in a traditional model-based context, relying on the model for likelihood-based methods, instead of the traditional PLS moment-based methods, to estimate the envelope in addition to other aspects of an analysis like standard errors. It also means that we can use envelopes to understand how PLS might contribute to a path analysis, both in its original moment-based version and the recent likelihood-based methods that arise from its connection with envelopes.  We develop that line of reasoning in this section, starting with an adaption of envelopes to path modeling.

\subsection{Envelope model}

Envelopes were introduced by \citet*{Cook2007} and developed as a means of increasing efficiency of estimation and prediction in multivariate linear regression by \citet{Cook2010}.  There is now a body of literature on envelope models and methods, including applications to reduced rank regression \citep{Cook:2015ab} and a theoretical basis for extending envelopes beyond multivariate linear models to almost any multivariate context \citep{CookZhang2015foundation}.  It is shown in this literature that envelope methods can result in efficiency gains equivalent to increasing the sample size many  times over.  The connections between envelopes and PLS regression established by \citet{CookHellandSu2013} are most relevant to this article.  We next adapt envelopes to the reflexive model of Section~\ref{setup} and then address the connection between this adaptation and PLS path modeling.

In reference to the reflexive model of Section~\ref{setup},  envelopes  gain efficiency by allowing for the possibilities that (a) there are {\em $\eta$-invariants}, which are defined as linear combinations of $Y$ whose distribution is unaffected by changes in $\eta$, or (b) there are {\em $\xi$-invariants}, which are defined as linear combinations of $X$ whose distribution is unaffected by changes in $\xi$, or (c) both (a) and (b) simultaneously.  An envelope, which is the novel population meta-parameter  estimated by moment-based PLS, is driven entirely by the possibility of $\eta$ and $\xi$-invariants.  Such quantities represent extraneous noise in the indicators that results in relatively inefficient estimators.  Envelopes are populations constructions that can be estimated by moment-based PLS and likelihood-based methods, both of which lead to  distinguishing variants from invariants to producing efficiency gains.

If the composite $G^{T}Y$ is $\eta$-invariant, then so is $AG^{T}Y$ for any nonsingular matrix $A$.  In consequence, a specific set of weights $G$ is not identifiable, but $\spn(G)$, the subspace spanned by the columns of $G$, is identifiable, which leads us to consider subspaces rather than individual coordinates in the formal incorporations of $\eta$ and $\xi$-invariants.

Let $\yspc \subseteq \real{r}$ and $\xspc \subseteq \real{p}$ be the smallest subspaces with the properties that 
\begin{eqnarray}
(i)\; Q_{\yspc}Y \mid \eta \sim Q_{\yspc}Y & \mathrm{\;and\;} & (ii) \; P_{\yspc}Y \indep Q_{\yspc}Y \mid \eta \label{condY} \\
(i) \; Q_{\xspc}X \mid \xi \sim Q_{\xspc}X & \mathrm{\;and\;} & (ii) \;P_{\xspc}X \indep Q_{\xspc}X \mid \xi, \label{condX}
\end{eqnarray}
where $P_{(\cdot)}$ is the projection onto the subspace indicated by its subscript, $Q_{(\cdot)} = I - P_{(\cdot)}$ is the projection onto the orthogonal complement, $\indep$ denotes independent variates and $\sim$ means equal in distribution.  The interpretation and implications of (\ref{condY}) and (\ref{condX}) are the same, only the variable and its latent construct change.  Because of this we focus on $Y$ and $\eta$ for a time, understanding that the same results and conclusions are applicable to $X$ and $\xi$.  Condition (\ref{condY}(i)) means that the marginal distribution of $Q_{\yspc}Y$ does not depend on $\eta$, while condition  (\ref{condY}(ii)) means that $Q_{\yspc}Y$ cannot furnish information about $\eta$ via an association with $P_{\yspc}Y$.  Together these two conditions imply that $Q_{\yspc}Y$ is the $\eta$-invariant part of $Y$ leaving $P_{\yspc}Y$ as the $\eta$-variant, the only part of $Y$ to be affected by $\eta$.  In reference to (\ref{model-2}) the subspace $\yspc$ is constructed formally as the smallest reducing subspace (See Appendix~\ref{sec:redsub}) of $\Sigma_{Y|\eta}$ that contains $\bspc_{Y|\eta} := \spn(\beta_{Y|\eta})$ and is denoted in full as  $\yspc_{\Sigma_{Y|\eta}}(\bspc_{Y|\eta})$, which is referred to as the $\Sigma_{Y|\eta}$-envelope of $\bspc_{Y|\eta}$ \citep{Cook2010}.  Applying these same ideas to $(X,\xi)$ we arrive at $\xspc_{\Sigma_{X|\xi}}(\bspc_{X|\xi})$, the $\Sigma_{X|\xi}$-envelope of $\bspc_{X|\xi} := \spn(\beta_{X|\xi})$. Let $u_{Y|\eta}$ and $u_{X|\xi}$ denote the dimensions of  $\yspc_{\Sigma_{Y|\eta}}(\bspc_{Y|\eta})$ and $\xspc_{\Sigma_{X|\xi}}(\bspc_{X|\xi})$.

These constructions result in envelope versions of models (\ref{model-2}) and (\ref{model-3}) as follows.  Let $\Gamma \in \real{r \times u_{Y|\eta}}$ and $\Phi \in \real{p \times u_{X|\xi}}$ denote semi-orthogonal basis matrices for $\yspc_{\Sigma_{Y|\eta}}(\bspc_{Y|\eta})$  and $\xspc_{\Sigma_{X|\xi}}(\bspc_{X|\xi})$.  Let $(\Gamma, \Gamma_{0}) \in \real{r \times r}$ and $(\Phi, \Phi_{0}) \in \real{p \times p}$ be orthogonal matrices.  Then
\begin{eqnarray}
 Y &=& \mu_{Y} + \Gamma \gamma (\eta -\mu_{\eta})+ \epsilon_{Y|\eta} ,  \mathrm{\;where\;} \epsilon_{Y|\eta} \sim N_{r}(0,\Gamma \Omega \Gamma^{T} + \Gamma_{0}\Omega_{0}\Gamma_{0}^{T}). \label{Yenvmodel}\\
 X & = & \mu_{X } +\Phi \phi(\xi-\mu_{\xi}) + \epsilon_{X|\xi} ,  \mathrm{\;where\;} \epsilon_{X|\xi} \sim N_{p}(0, \Phi \Delta \Phi^{T} + \Phi_{0}\Delta_{0}\Phi_{0}^{T}). \label{Xenvmodel}
\end{eqnarray}
Here $\gamma \in \real{u_{Y | \eta}}$ and $\phi \in \real{u_{X|\xi}}$ give the coordinates of $\beta_{Y|\eta}$ and $\beta_{X|\xi}$ in terms of the basis matrices $\Gamma$ and $\Phi$, and $\Delta$, $\Delta_{0}$, $\Omega$ and $\Omega_{0}$ are positive definite matrices with dimensions that conform to the indicated matrix multiplications.  To see how this structure reflects the $\eta$-variant and $\eta$-invariant parts of $Y$, multiply $Y$ by the orthogonal matrix $(\Gamma, \Gamma_{0})^{T}$ to get a transformed representation in terms of composite indicators,
\begin{eqnarray*}
 \Gamma^{T}Y &=& \Gamma^{T}\mu_{Y} + \gamma (\eta -\mu_{\eta})+ \Gamma^{T}\epsilon_{Y|\eta} ,  \mathrm{\;where\;} \Gamma^{T}\epsilon_{Y|\eta} \sim N_{u_{Y|\eta}}(0, \Omega) \mathrm{\;with\;}\Omega = \Sigma_{\Gamma^{T}Y|\eta}\\
 \Gamma_{0}^{T}Y &=& \Gamma_{0}^{T}\mu_{Y} +\Gamma_{0}^{T} \epsilon_{Y|\eta} ,  \mathrm{\;where\;} \Gamma_{0}^{T}\epsilon_{Y|\eta} \sim N_{r}(0,\Omega_{0}) \mathrm{\;with\;} \Omega_{0} = \Sigma_{\Gamma_{0}^{T}Y|\eta}. 
\end{eqnarray*}
Since the marginal distribution of $\Gamma_{0}^{T}Y$ does  not depend on $\eta$ and $\Gamma_{0}^{T} \epsilon_{Y|\eta} \indep \Gamma^{T} \epsilon_{Y|\eta}$ we see that only the composite indicator $\Gamma^{T}Y$ reflects the latent construct $\eta$.  In this context, envelopes gain efficiency by identifying and accounting for  $\Gamma_{0}^{T}Y$, the $\eta$-invariant part of $Y$.  The increase in efficiency can be massive if variation in the $\eta$-invariant components of $\Gamma_{0}^{T}Y$  is substantially larger than the variation of the $\eta$-variant components of $\Gamma^{T}Y$.  Informally, envelopes will result in substantial efficiency gains if the indicator $Y$ contains non-trivial noise that is unrelated to the latent construct $\eta$.  The weights $\Gamma$ used in forming the composite indicator $\Gamma^{T}Y$ are not identifiable, but $\spn(\Gamma) = \yspc_{\Sigma_{Y|\eta}}(\bspc_{Y|\eta})$ is identifiable and this is sufficient to allow improved estimation of $\cor \{E(\eta \mid Y), E(\xi \mid X)\}$.

It follows from the envelope model given by (\ref{Yenvmodel}) and (\ref{Xenvmodel}) that 
\begin{eqnarray*}
\Sigma_{Y} & = & \Gamma (\gamma\gamma^{T}\sigma_{\eta}^{2} + \Omega) \Gamma^{T} + \Gamma_{0}\Omega_{0}\Gamma_{0}^{T} \nonumber \\
\Sigma_{YX} & = & \Gamma \gamma \phi^{T}\Phi^{T} \sigma_{\xi \eta} \nonumber  \\
\Sigma_{X} & = & \Phi(\phi\phi^{T}\sigma_{\xi}^{2} + \Delta)\Phi^{T} + \Phi_{0}\Delta_{0}\Phi^{T}\nonumber  \\
|\cor \{E( \xi|X),E(\eta|Y)\}|  
 & = & |\cor \{E( \xi|\Phi^{T}X),E(\eta|\Gamma^{T}Y)\}|.
\end{eqnarray*}
The essential implication of this last result is as follows.  Once estimators $\Gammahat$ and $\Phihat$ are known (see Section~\ref{sec:basisest}), we can use the estimated composite indicators  $\Gammahat^{T}Y$ and $\Phihat^{T}X$ in place of $Y$ and $X$ in (\ref{est}) to estimate  $|\cor (E( \xi|X),E(\eta|Y))| $.  

A similar connection between envelopes and $\cor(\eta,\xi)$ is problematic.  Since only $\spn(\Gamma) = \yspc_{\Sigma_{Y|\eta}}(\bspc_{Y|\eta})$ and $\spn(\Phi)=\xspc_{\Sigma_{X|\xi}}(\bspc_{X|\xi})$ are identifiable, arbitrary choices of bases for these subspaces will not likely result in $\Sigma_{\Gamma^{T}Y | \eta} = \Omega$ and $\Sigma_{\Phi^{T}X|\xi}=\Delta$ being diagonal matrices, as required by SEM.
Choosing arbitrary bases $\Gamma$ and $\Phi$ and using the marginal constraints, the envelope composites have the following structure,
\begin{eqnarray*}
\Sigma_{\Gamma^{T}Y} & = &  \gamma\gamma^{T} + \Omega \nonumber \\
\Sigma_{\Gamma^{T}Y,\Phi^{T}X} & = & \gamma \phi^{T} \cor(\xi, \eta) \nonumber  \\
\Sigma_{\Phi^{T}X} & = & \phi\phi^{T} + \Delta. \nonumber  
\end{eqnarray*}
From this we see that the joint distribution of the envelope composites $(\Gamma^{T}Y, \Phi^{T}X)$ has the same structure as SEM shown in Appendix~\ref{sec:proofs}, equation (\ref{SIGMA2}), except that assuming $\Omega$ and $\Delta$ to be diagonal matrices is untenable from this structure alone.  Additionally, $\Omega$ and $\Delta$ are not identifiable because they are confounded with $\gamma \gamma^{T}$ and $\phi \phi^{T}$.

In short, PLS is founded on the possibility that there are $\eta$ and $\xi$-invariants, while SEM is founded on the condition that $\Sigma_{Y|\eta}$ and $\Sigma_{X|\xi}$ be diagonal matrices.  It is not now clear how to resolve the conflict between these two starting points and thus it is also unclear if PLS has any role in standard SEM for estimating $|\cor(\eta, \xi)|$ directly.

\subsection{Estimating bases $\Gamma$ and $\Phi$ }\label{sec:basisest}

The role of moment-based or likelihood-based PLS in path analysis is to provide an estimator of $|\cor\{E( \xi|X),E(\eta|Y)\}|$ that is less variable than the estimator given in (\ref{est}) without  requiring $\Sigma_{X|\xi}$ or $\Sigma_{Y|\eta}$ to be diagonal matrices as in SEM.  
The essential ingredient supplied by moment-based PLS is an algorithm for estimating bases $\Gammahat$ and $\Phihat$ of $\spn(\Gamma)=\yspc_{\Sigma_{Y|\eta}}(\bspc_{Y|\eta})$ and $\spn(\Phi)=\xspc_{\Sigma_{X|\xi}}(\bspc_{X|\xi})$ from the multivariate regressions of $Y$ on $X$ and $X$ on $Y$ \citep{CookHellandSu2013}.  The fundamental estimation problem addressed by PLS is then to estimate $\yspc_{\Sigma_{Y|\eta}}(\bspc_{Y|\eta})$ and $\xspc_{\Sigma_{X|\xi}}(\bspc_{X|\xi})$ \citep[See][Section 2.1, paragraph 1]{Ronkko2016}.  While there are variations in how PLS algorithms alternate between these two regressions, the basic algorithm is as developed by \citet{deJong1993}. See also \citet[][Section 3]{Tenenhaus2005}.  

More directly, \citet*{CookZhang2015foundation} developed a likelihood-based methodology for estimating $\yspc_{\Sigma_{Y|\eta}}(\bspc_{Y|\eta})$ and $\xspc_{\Sigma_{X|\xi}}(\bspc_{X|\xi})$, assuming that $X$ and $Y$ jointly follow a multivariate normal distribution. They showed that moment-based PLS estimators of the parameters in the distribution of $(X^{T},Y^{T})$ can be  dominated by the likelihood-based estimator from their simultaneous envelope model, and that without normality their estimators are still $\sqrt{n}$-consistent. While this likelihood-based methodology can be adapted straightforwardly to path analyses based on the reflexive model of Section~\ref{setup}, it does not make use of the fact that the coefficient matrix has rank 1, as we described in Section~\ref{RRR}.  Following \citet{Cook:2015ab}, these two approaches can be combined, leading to a new simultaneous envelope reduced rank (SERR) estimator.

In moment-based PLS path analysis, its maximum likelihood counterpart by  \citet*{CookZhang2015foundation} and SERR, the dimensions $u_{Y|\eta}$ and $u_{X|\xi}$ are not specified in advance but are estimated along with the other parameters.  In consequence, the covariance structures of the measurement errors $\Sigma_{Y|\eta}$ and $\Sigma_{X|\xi}$ shown (\ref{Yenvmodel}) and (\ref{Xenvmodel}) are not assumed but are determined by the data from within the class of $(r+1)(p+1)$ covariance structure obtained by varying $u_{Y|\eta} = 0,\ldots,r$ and $u_{X|\xi}=0,\ldots,p$.  If there are no $\eta$ or $\xi$-invariants then $u_{Y|\eta}=r$ and $u_{X|\xi}=p$, and models (\ref{Yenvmodel}) and (\ref{Xenvmodel}) reduce to the starting models shown in (\ref{model-2}) and (\ref{model-3}). On the other extreme, if $u_{Y|\eta}=0$ and $u_{X|\xi}=0$ then the indicators are independent of the latent constructs.

\section{Simulation example of \citet{Ronkko2016}}\label{sec:example}

\begin{figure}[ht]
\begin{center}
$N=100$ \hspace*{6cm}  $N=1000$

\includegraphics[width=0.48\textwidth]{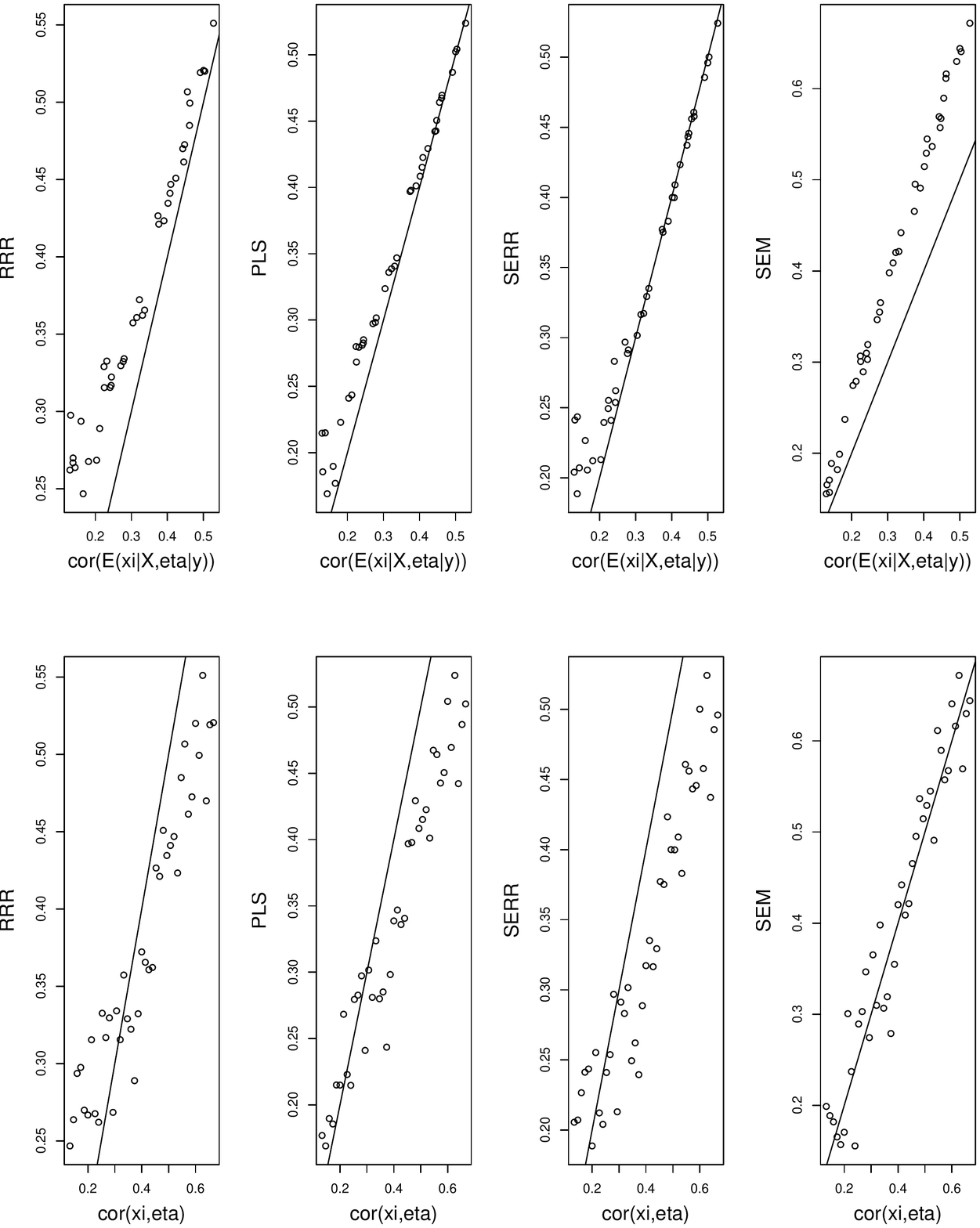}  \hspace*{0.3cm} \includegraphics[width=0.48\textwidth]{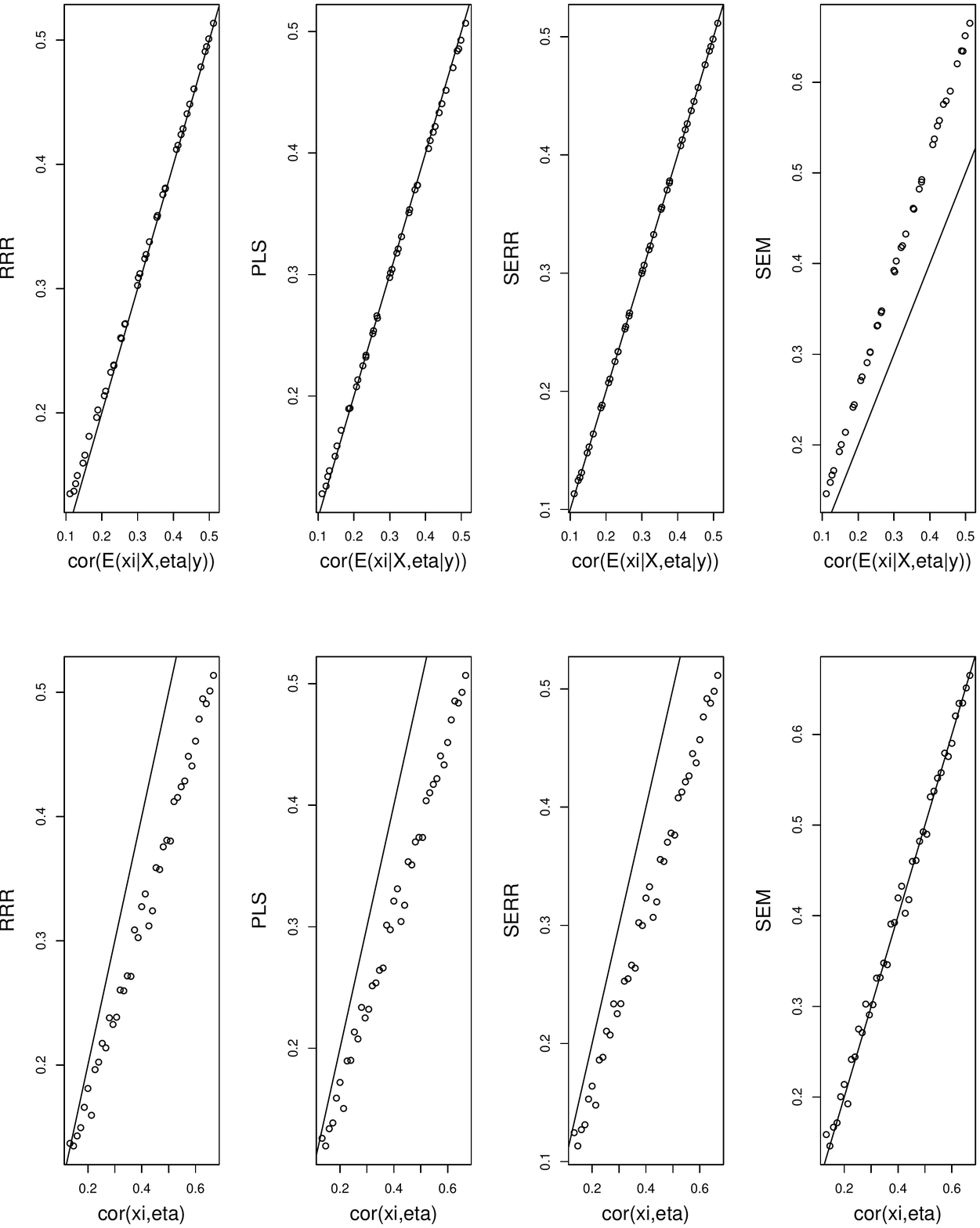}
\caption{$L=  (4/3, 0.98, 3/4)^{T}$, $\Sigma_{X|\xi}=\Sigma_{Y|\eta}=I_3$.  pls refers to moment-based PLS.}  \label{fig1}
\end{center}
\vspace{-0.4cm}
\end{figure}

 \citet{Ronkko2016} illustrated some of their concerns via simulations with
$p=r=3$, sample size $N=100$,
$\mu_{\xi} =\mu_{\eta} =0$,
 $\sigma_{\xi}=\sigma_{\eta}=1$, $ \sigma_{\xi \eta}= \cor(\eta, \xi) \in (0,2/3)$,
$\mu_{X}= \mu_{Y} =0$,
$ \beta_{Y|\eta} =\beta_{X|\xi}=L=  (4/3, 0.98 ,3/4)^T$ and
$\Sigma_{X| \xi}=\Sigma_{Y|\eta}=I$.
It follows from this setup that
\begin{eqnarray*}
\cor \{E(\eta|Y), E(\xi|X)\} &=& \cor \{E(\etat|Y), E(\xit|X)\}  \nonumber \\
& = & L^T L(I+L^TL)^{-1}\cor(\eta, \xi) = 0.77 \cor(\eta,\xi) \leq \cor(\eta, \xi).
\end{eqnarray*}
We simulated $N=100$ and $N=1000$ observations on $(X,Y)$ according to these settings with various values for $\cor(\eta, \xi)$ and then applied RRR, PLS, SEM and the simultaneous envelope reduced-rank estimator (SERR) based on the methods of \citet{CookZhang2015} and \citet{Cook:2015ab}.  We can see from the results shown in  Figure \ref{fig1}, which are the averages over 10 replications (\citet{Ronkko2016} used one replication), that
at $N = 1000$ SEM does well estimating  $\cor (\xi, \beta)$, while RR,  PLS and SERR  do well estimating $\cor \{ E(\eta|Y), E(\xi|X)\}$.  At $N=100$, RRR clearly overestimates $\cor \{E(\eta|Y), E(\xi|X)\} $ for all $\cor(\eta, \xi) \in (0, 2/3)$.  There is also a tendency for envelopes and SERR to overestimate $\cor \{E(\eta|Y), E(\xi|X)\} $ at smaller values of $\cor \{E(\eta|Y), E(\xi|X)\} $, with envelopes performing a bit better.  This arises because for small correlations the signal strength is weak, making the weights $\Gamma$ and $\Phi$ difficult to estimate.  The predicted relationship that $|\cor \{E(\eta|Y), E(\xi|X)\}| \leq |\cor(\xi, \eta)|$ is also demonstrated in Figure~\ref{fig1}.   Appendix~\ref{sec:addexamples} gives results for relatively weak signals (more bias) with $L=(0.58,0.98,0)^T$ and for relatively strong signals (less bias) with $L=(4,4,4)^T$.

\begin{figure}[ht]
\begin{center}
$N=100$ \hspace*{6cm}  $N=1000$

\includegraphics[width=0.48\textwidth]{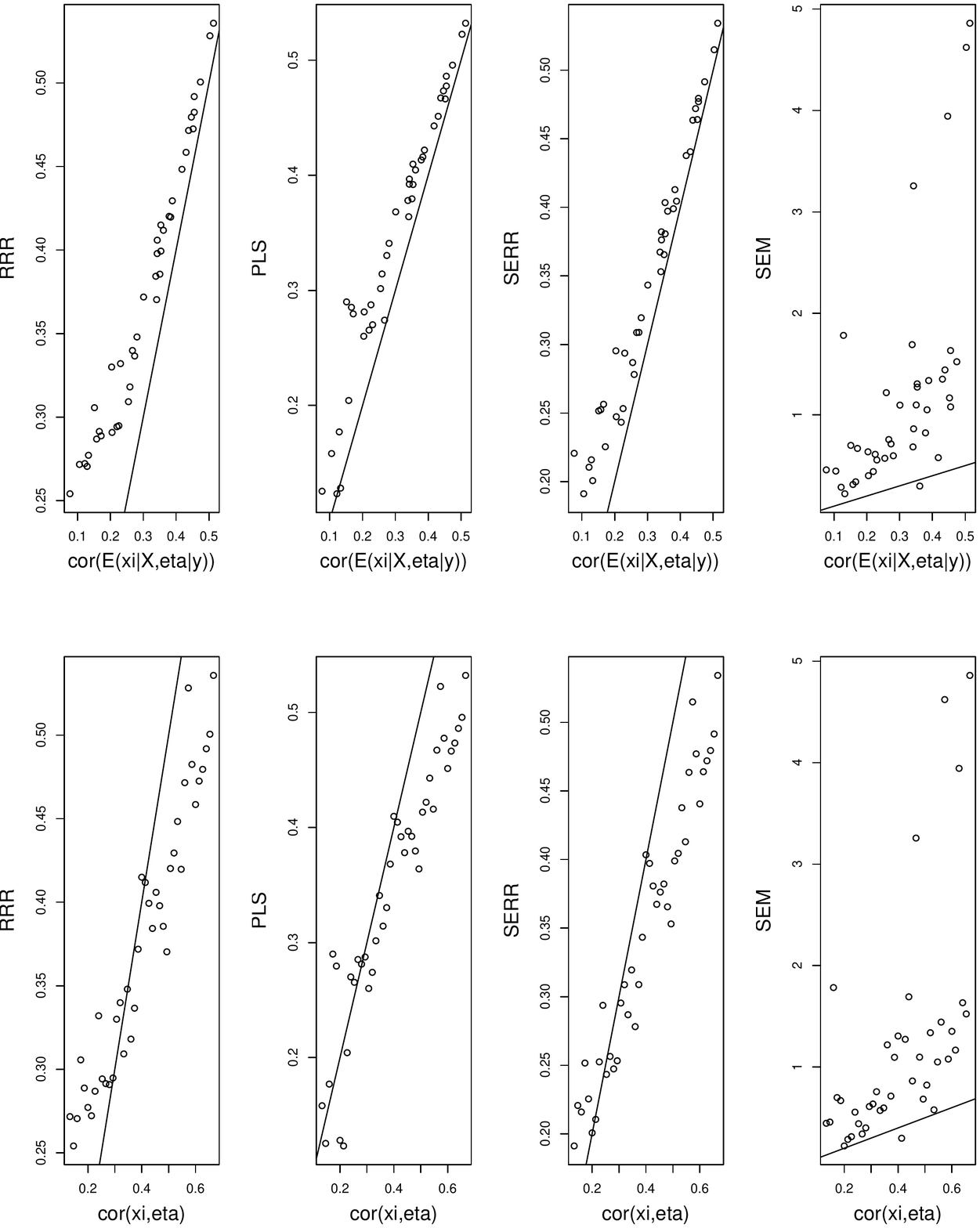}  \hspace*{0.3cm} \includegraphics[width=0.48\textwidth]{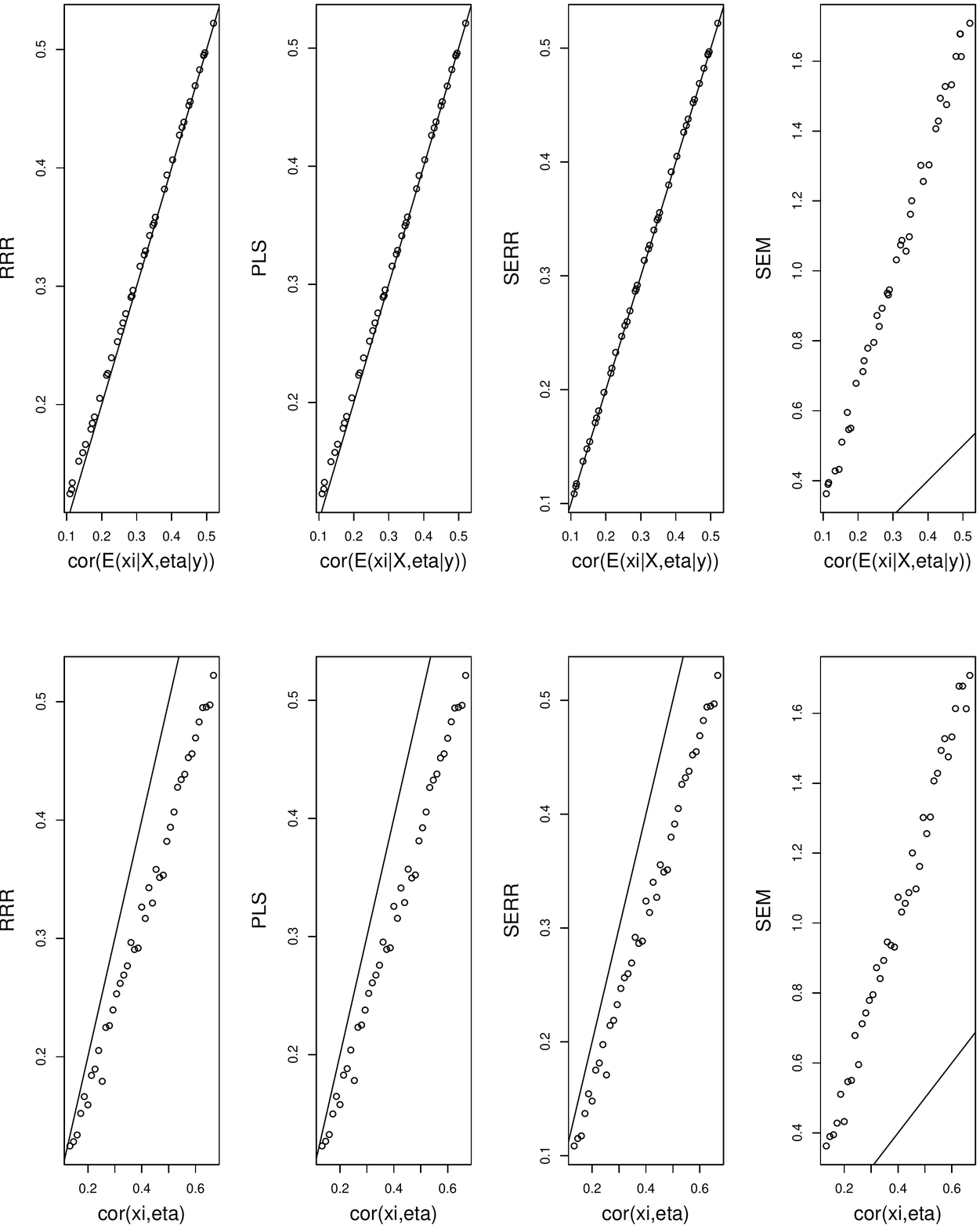}

\caption{$L=  (4/3 ,0.98, 3/4)^{T}$, $\Sigma_{X|\xi}=\Sigma_{Y|\eta} = L (L^TL)^{-1}L^T + 3 L_0 L_0^T$}\label{fig3}
\end{center}
\vspace{-0.4cm}
\end{figure}

Simulating with  $\Sigma_{X|\xi} =\Sigma_{Y|\eta} =  I$, as we did in the construction of Figure~\ref{fig1},  conforms to the SEM assumption that these matrices be diagonal and partially accounts for the success of SEM in Figure~\ref{fig1}.  To illustrate the importance of these assumptions, we next took   $\Sigma_{X|\xi}=\Sigma_{Y|\eta} = L (L^TL)^{-1}L^T + 3 L_0 L_0^T$
with $L_0 \in {\mathbb R}^{3 \times 2}$ constructed as a semi-orthogonal complement of $L$. With this structure $L_{0}^{T}Y$ and $L_{0}^{T}X$ are $\eta$ and $\xi$-invariants which satisfy the structural assumptions for moment-based PLS and envelopes.  The results with $L = (4/3, 0.98, 3/4)$ for RRR, PLS and SERR shown in Figure \ref{fig3} are qualitatively same as those shown in Figure~\ref{fig1}.   However, now SEM  apparently fails to give a useful estimate of either $\cor \{E(\eta|Y), E(\xi|X)\} $ or $\cor(\eta, \xi)$.

\section{Myths and urban legends}\label{sec:myths}

We are now in a position to provide a commentary that addresses some of  the myths and urban legends \citep{Ronkko2016} surrounding PLS in path modeling.  In doing so, we reiterate the main points of our previous discussion.

\paragraph{PLS versus SEM.} The fundamental difference between PLS and SEM rests with the assumptions and goals.  Traditional moment-based PLS, envelopes and the SERR implementation we used for illustration all  rely on the presence of $\eta$ and $\xi$-invariants to produce an efficient estimator of $|\cor\{E( \xi|X),E(\eta|Y)\}|$  without requiring $\Sigma_{Y|\eta}$ and $\Sigma_{X|\xi}$ to be diagonal matrices.  One goal of SEM is to estimate $|\cor(\eta,\xi)|$ while requiring that $\Sigma_{Y|\eta}$ and $\Sigma_{X|\xi}$ be diagonal matrices.  The current literature does not seem to allow straightforward resolution of these conflicting structures.

\paragraph{PLS is SEM?}  We do not find PLS in its moment-based implementations to be an instance of structural equation modeling.  We understand that maximum likelihood estimation is one requirement for a method to be dubbed SEM.  Since current PLS methods are moment-based they do not meet this requirement.  However, as mentioned previously, the new simultaneous envelope model of \citet{CookZhang2015} uses maximum likelihood to estimate bases for $\yspc_{\Sigma_{Y|\eta}}(\bspc_{Y|\eta})$  and $\xspc_{\Sigma_{X|\xi}}(\bspc_{X|\xi})$.  Since in this application envelope methodology can be viewed as a maximum likelihood version of PLS, it does qualify as an SEM for estimating $|\cor\{E( \xi|X),E(\eta|Y)\}|$, but not for estimating $|\cor(\eta,\xi)|$.

\paragraph{What do the PLS weights actually accomplish?}  Estimates of the true PLS weights $\Gamma$ and $\Phi$ are intended to extract the $\eta$-variant and $\xi$-variant parts of the indicators $Y$ and $X$, giving the composites $\Gamma^{T}Y$ and $\Phi^{T}X$ and effectively leaving behind the $\eta$-invariant and $\xi$-invariant noise represented by $\Gamma_{0}^{T}Y$ and $\Phi_{0}^{T}X$.  The estimated weights may provide important insights into how the latent constructs $\eta$ and $\xi$ influence the indicators $Y$ and $X$, but here it should be remembered that the weights are not themselves identifiable but their spans $\yspc_{\Sigma_{Y|\eta}}(\bspc_{Y|\eta})$ and $\xspc_{\Sigma_{X|\xi}}(\bspc_{X|\xi})$ are identifiable.  This means that any orthogonal  transformation $O$ of an estimated set of weights, $\Gammahat \mapsto \Gammahat O$, produces an equivalent solution, but the apparent interpretation of the two sets of weights $\Gammahat$ and $ \Gammahat O$ may seem quite different.  This may partially address the concern expressed by \citet[][Section 2.2, penultimate paragraph]{Ronkko2016} over model-dependent weights.

If there are no $\eta$ or $\xi$-invariants, then all linear combinations of $Y$ and $X$ respond to changes in $\eta$ and $\xi$, an inference that might be useful in some studies.  In such cases we do not now see that PLS has much to offer, although it is conceivable that it could still offer advantages in mean squared error.

\paragraph{Are PLS weights optimal?}  This cannot be answered without a clear statement of the optimality criterion.  Moment-based PLS regression produces $\sqrt{n}$-consistent estimators and on those grounds it clears a first hurdle for a reasonable statistical method.  The simultaneous envelope model of \citet{CookZhang2015} uses maximum likelihood to estimate bases for $\yspc_{\Sigma_{Y|\eta}}(\bspc_{Y|\eta})$  and $\xspc_{\Sigma_{X|\xi}}(\bspc_{X|\xi})$ and thus inherits its optimality properties from general likelihood theory under normality.  In the absence if normality it can also produce $\sqrt{n}$-consistent estimators.

\paragraph{Do PLS weights reduce the impact of measurement error?}  Yes.  By accounting for the variant and invariant parts of $Y$ and $X$, PLS effectively reduces the measurement errors by amounts that depends on $\Omega_{0} = \var(\Gamma_{0}^{T}Y)$ and $\Delta_{0}=\var(\Phi_{0}^{T}Y)$.

\paragraph{PLS versus unit-weighted summed composites.}  Let $1_{q}$ denote a $q \times 1$ vector of ones.  Then unit-weighted summed composites are $Z_{Y}=1_{r}^{T}Y$ and $Z_{X} = 1_{p}^{T}X$.  The effectiveness of $Z_{Y}$ and $Z_{X}$ depends on the relationship between $\spn(1_{r})$ and $\yspc_{\Sigma_{Y|\eta}}(\bspc_{Y|\eta})$  and between $\spn(1_{p})$ and $\xspc_{\Sigma_{X|\xi}}(\bspc_{X|\xi})$.  If $\spn(1_{r}) \subseteq \yspc_{\Sigma_{Y|\eta}}(\bspc_{Y|\eta})$ then we might expect $Z_{Y}$ to be a useful composite depending on the size of $u_{Y|\eta}$.  On the other extreme, if $\spn(1_{r}) \subseteq \yspc^{\perp}_{\Sigma_{Y|\eta}}(\bspc_{Y|\eta})$ then $Z_{Y}$ is an $\eta$-invariant composite and no useful results should be expected.  In short, the usefulness of summed composites depends on the anatomy of the paths.  In some path analyses they might be quite useful, while useless in others.

\paragraph{Does PLS have advantages in small samples?}  An assessment of this issue depends on the meaning of ``small.''   If the sample size $n$ is large enough to ensure that $\Sigma_{X}$, $\Sigma_{Y}$ and $\Sigma_{YX}$ are well-estimated by their sample versions then we see no particular advantage to PLS on the grounds of the sample size.  On the other extreme, if $n < r$ and $n < p$ then the sample versions of $\Sigma_{X}$, $\Sigma_{Y}$ and $\Sigma_{YX}$ have less than full rank and consequently maximum likelihood (including SEM) is not generally a serviceable criterion.  Here moment-based PLS may have an advantage as it does not require that  the sample versions of $\Sigma_{X}$ and $\Sigma_{Y}$ be positive definite.  For instance,  \citet{CookForzani2017PLS, CookForzani2018} showed that in high-dimensional univariate regressions moment-based PLS predictions can converge at the $\sqrt{n}$ rate regardless of the relationship between $n$ and the number of predictors.

\paragraph{Bias.}  If the goal is to estimate $|\cor(\eta,\xi)|$ with PLS then bias is a relevant issue.  But if the goal is to estimate $|\cor\{E( \xi|X),E(\eta|Y)\}|$ then bias, as we understand its targeted meaning in the path modeling literature, is no longer relevant.  It follows from our discussion of bias in Section~\ref{bias} that $|\cor\{E( \xi|X),E(\eta|Y)\}| \leq |\cor(\eta,\xi)|$.  When the goal is to estimate $|\cor(\eta,\xi)|$, it may be useful to have knowledge of the lower bound $|\cor\{E( \xi|X),E(\eta|Y)\}| $ that does not require restrictive conditions on $\Sigma_{Y|\eta}$ and $\Sigma_{X|\xi}$. 

\paragraph{Principal components.}  \citet{Ronkko2016}, as well as other authors, mention the possibility of using weighting schemes other than PLS to construct composites.  Here we discuss principal components as an instance of such alternatives.  \citet{TippingBishop1999} provided a latent variable model that yields principal components via maximum likelihood estimation.  Their model adapted to the present context yields models (\ref{model-2}) and (\ref{model-3}) with the important additional restriction that the measurement errors be isotropic: $\Sigma_{Y|\eta} = \sigma_{Y|\eta}^{2}I_{r}$ and $\Sigma_{X|\xi} = \sigma_{X|\xi}^{2}I_{p}$, where $I_{q}$  denotes the identity matrix of dimension $q$.  This restricts the individual indicators to be uncorrelated conditional on the latent variables, as in SEM, and to have the same conditional variances.  With these restrictions, Tipping and Bishop showed that, under normality, the maximum likelihood estimators  of $\spn(\beta_{Y|\eta})$ and $\spn(\beta_{X|\xi})$ are the spans of the first eigenvectors $\ell_{Y}$ and $\ell_{X}$ of $\Sigmahat_{Y}$ and $\Sigmahat_{X}$, yielding the composites $\ell_{Y}^{T}Y$ and $\ell_{X}^{T}X$.  If isotropic measurement errors are accepted then principal components is a tenable method.

\paragraph{Composite factor models.}   As mentioned in Section~\ref{sec:intro}, the composite factor model described by \citet[][p. 216]{Mcintosh2014} takes the constructs to be exact linear combinations of the indicators, $\eta = \beta_{\eta}^{T}Y$, $\xi = \beta_{\xi}^{T}X$ with the requirement that $Y \indep X \mid \beta_{\eta}^{T}Y, \beta_{\xi}^{T}X$.  As a consequence of this structure, (a) the marginal and regression constraints both reduce to $\beta_{\xi}^{T}\Sigma_{X}\beta_{\xi} =  \beta_{\eta}^{T}\Sigma_{Y}\beta_{\eta} = 1$ since $E(\xi \mid X) = \xi$ and $E(\eta \mid Y) = \eta$, and (b) the measures of association are the same, $\cor\{E(\xi\mid X), E(\eta \mid Y)\} = \cor(\xi, \eta) = \beta_{\xi}^{T}\Sigma_{XY}\beta_{\eta}$.   Since $(X,Y)$ is multivariate normal, $Y \indep X \mid \beta_{Y|X}X$ and $Y \indep X \mid \beta_{X|Y}Y$, which implies that $Y \indep X \mid (\beta_{Y|X}X, \beta_{X|Y}Y)$.  In consequence, we must have $\rank(\beta_{Y|X})  = \rank(\beta_{X|Y}) = 1$, $\spn(\beta_{Y|X}) = \spn(\beta_{\xi}^{T})$ and  $\spn(\beta_{X|Y}) = \spn(\beta_{\eta}^{T})$.  This structure means that $\cor(\xi, \eta)$ can be estimated consistently from the RRR model discussed in Section~\ref{sec:RRR} or from its envelope version described in Section~\ref{sec:PLS}.

The essential point here is that by modifying the definition of the constructs, most of the myths and urban legends described by \citet{Ronkko2016} no longer seem to be at issue.  Of course this requires  that the modified definition of the constructs be useful in the first place.

\section*{Acknowledgements} We thank Marilina Carena for help drawing Figure~\ref{figpath} and Marko Sarstedt for helpful comments on an early draft.

\appendix
\section{Reducing subspaces}\label{sec:redsub}

\begin{definition}\label{def:reducingsubspace} A subspace $\rspc\subseteq \real{r}$
is said to be a reducing subspace of the real symmetric $r \times r$ matrix $M$ if
$\rspc$ decomposes $\M$ as $\M=\Pbf_{\rspc}\M\Pbf_{\rspc}+\Q_{\rspc}\M\Q_{\rspc}$.
If $\rspc$ is a reducing subspace of $\M$, we say that $\rspc$
reduces $\M$.
\end{definition}

This definition of a reducing subspace is equivalent to that used
by \citet{Cook2010}. It is common in the literature on invariant subspaces and functional analysis, although the underlying
notion of ``reduction'' differs from the usual understanding in statistics.  Here it is used to guarantee conditions (\ref{condY}) and (\ref{condX}).  The following definition makes use of reducing subspaces.

\begin{definition}\label{def:envelope} Let $\M$ be a real symmetric $r \times r$ matrix, and let $\bspc\subseteq\spn(\M)$. Then the $\M$-envelope of
$\bspc$, denoted by $\espc_{\M}(\bspc)$, is the intersection of
all reducing subspaces of $\M$ that contain $\bspc$.
\end{definition}

\section{Proofs and supporting results}\label{sec:proofs}

\begin{lemma}\label{lemma-one}
For model presented in Section \ref{setup} we have
\[ \left( \begin{array}{c}X \\ Y \\ \xi \\ \eta \end{array} \right)
\sim  \left( \begin{array}{c}\mu_{X} \\  \mu_{Y} \\\mu_{\xi}  \\ \mu_{\eta} \end{array} \right) + \epsilon_{X,Y,\xi,\eta}\]
where  $\var ( \epsilon_{X,Y,\xi,\eta})=\Sigma_{X,Y,\xi,\eta}$ with
\begin{equation}\label{SIGMA} \Sigma_{X,Y,\xi,\eta} = \left(\begin{array}{cccc} \Sigma_{X\mid \xi} + \beta_{X\mid \xi} \beta_{X\mid \xi}^T \sigma^2_{\xi}  & \beta_{X\mid \xi}\beta_{Y\mid \eta}^T \sigma_{\xi \eta} &\beta_{X\mid \xi}  \sigma^2_{\xi}   &  \beta_{X\mid \xi} \sigma_{\xi \eta}\\
\dots&  \Sigma_{Y\mid \eta} +\beta_{Y\mid \eta}    \beta_{Y\mid \eta}^T \sigma^2_{\eta}&\beta_{Y\mid \eta}  \sigma_{\xi \eta}  & \beta_{Y\mid \eta}  \sigma^2_{\eta}  \\
\dots & \dots & \sigma^2_{\xi}   & \sigma_{\xi \eta} \\
\dots & \dots & \dots & \sigma^2_{\eta}
\end{array}\right). \end{equation}
\end{lemma}

\proof
 The first two elements of the diagonal are direct consequence of the fact that $\cov (Z) = E(\cov (Z|H)) + \var (E (Z|H))$. For $\Sigma_{X\xi} $ and $\Sigma_{Y\eta}$ we use the fact that $\beta_{X|\xi} = \sigma_{\xi}^{-2} \Sigma_{X\xi}$ and $\beta_{Y|\eta} = \sigma_{\eta}^{-2} \Sigma_{Y\eta}$. To compute $\Sigma_{Y \eta}$ we use that
\begin{eqnarray*}
\Sigma_{Y \eta} &=& E( (Y-\mu_Y)(\eta-\mu_\eta))\\
&=&E_\eta [ E( (Y-\mu_Y)(\eta-\mu_\eta))|\eta]\\
&=&  \beta_{Y|\eta} E_\eta (\eta-\mu_\eta)^2\\
&=&  \beta_{Y|\eta} \sigma_{\eta}^{2}
\end{eqnarray*}
Analogously for $\Sigma_{X\xi}$. Now for $\Sigma_{XY} $ we use that
\begin{eqnarray*}
\Sigma_{Y X} &=& E ( (Y-\mu_Y)(X-\mu_X)^T)\\
&=& E_{\eta,\xi} [E_{Y,X|(\eta,\xi)}\{(Y-\mu_Y)(X-\mu_X)^T\}]\\
&=&\beta_{Y|\eta}  E_{\eta,\xi} (\eta-\mu_\eta)(\xi - \mu_{\xi}) \beta_{X|\xi}^T\\
&=&\beta_{Y|\eta} \sigma_{\xi \eta} \beta_{X|\xi}^T
\end{eqnarray*}
Now, for  $\Sigma_{X \eta}$ 
\begin{eqnarray*}
\Sigma_{X \eta} &=& E( (X-\mu_X)(\eta- \mu_\eta))\\
&=& E_{\eta,\xi} E ( (X-\mu_X)(\eta- \mu_\eta)|\eta,\xi)\\
&=&\beta_{X\mid \xi} E_{\eta,\xi} E ((\xi - \mu_\xi) (\eta- \mu_\eta)|\eta,\xi)\\
&=&\beta_{X\mid \xi} \sigma_{\xi \eta}
\end{eqnarray*}
The proof of $\Sigma_{Y \xi}$ is analogous.\eop

\begin{prooflem}{RRR}
This is direct consequence of Lemma \ref{lemma-one} since 
$\Sigma_{YX} = \beta_{Y|\eta} \sigma_{\xi \eta} \beta_{X|\xi}^T =
\Sigma_{Y\eta} \sigma_{\eta}^{-2} \sigma_{\xi \eta} \sigma_{\xi}^{-2} \Sigma_{\xi X}$
and therefore rank of $\Sigma_{YX} $ is one and the lemma follows. \end{prooflem}

\begin{proofprop}{one}
The first part follows from reduced rank model literature (see for example \citet*{Cook:2015ab}). Now,  using (\ref{SIGMA}) and the fact that $\mu_\eta=0$,
\begin{eqnarray*}
E (\eta|Y) &=& \Sigma_{\eta Y} \Sigma_Y^{-1} (Y- \mu_Y)
\end{eqnarray*}
and therefore 
\begin{eqnarray} \var (E(\eta|Y)) &=& \Sigma_{\eta Y} \Sigma_Y^{-1} \Sigma^T_{\eta Y} = 1 \label{mm}, \end{eqnarray}
where we use the hypothesis that $\var (E(\eta|Y))=1 $ in the last equal.
In the same way
\begin{equation}\label{mm1}
 \var (E(\xi|X))=\Sigma_{\xi X} \Sigma_X^{-1} \Sigma^T_{\xi X}=1.
 \end{equation}
Now, since
\begin{eqnarray}
 \Sigma_{YX} &=& A B^T  = \Sigma_{Y\eta} \sigma_{\eta}^{-2} \sigma_{\xi \eta} \sigma_{\xi}^{-2} \Sigma_{\xi X}\label{cuatros} \end{eqnarray}
we have for some $m_\eta$ and $m_\xi$ that
\begin{eqnarray}
\Sigma_{Y \eta} &=& A m_\eta \label{meta}\\
\Sigma_{X \xi} &=& B m_\xi \label{mxi}
\end{eqnarray}
(\ref{cuatros}), (\ref{meta}) and (\ref{mxi}) together makes
$ A B^T = A m_\eta   \sigma_{\eta}^{-2} \sigma_{\xi \eta} \sigma_{\xi}^{-2}  m_\xi B^T$
 and therefore 
\begin{eqnarray}\label{new1}
  m_\eta \sigma_{\eta}^{-2} \sigma_{\xi \eta} \sigma_{\xi}^{-2} m_\xi & =&1
  \end{eqnarray}

Now, using (\ref{mm}) and (\ref{mm1}) together with (\ref{meta}) and (\ref{mxi}) we have that
\begin{eqnarray}
m_\eta^2 & = & (A^T \Sigma_{Y}^{-1} A)^{-1} \label{meta1}\\
m_\xi^2 &=& (B^T \Sigma_{X}^{-1} B)^{-1} \label{mxi1}
\end{eqnarray} 

Plugging this into (\ref{new1}),
\begin{eqnarray*}
 \sigma_{\eta}^{-2} \sigma_{\xi \eta} \sigma_{\xi}^{-2}&=& (m_\xi m_\eta)^{-1} 
 \\
 &=& (A^T \Sigma_Y^{-1} A)^{1/2} (B^T \Sigma_X^{-1} B)^{1/2}\\
 & = & (A^T \Sigma_Y^{-1} AB^T \Sigma_X^{-1} B)^{1/2}  \\
& = & \tr^{1/2}( (AB^{T})^{T} \Sigma_Y^{-1} AB^T \Sigma_X^{-1})  \\
& = & \tr^{1/2}(\beta_{Y|X}\beta_{X|Y}).
   \end{eqnarray*}
%
  Let us note that from (\ref{meta1}) and (\ref{mxi1}), $m_\xi^2$ and $m_\eta^2$ are not unique, nevertheless $m_\eta^2 m_\xi^2$ is unique since any change of 
  $A$ and $B$ should be such $AB^T$ is the same. As a consequence
  $m_\eta  m_\xi $ is unique except for a sign. And therefore $ \sigma_{\eta}^{-2} \sigma_{\xi \eta} \sigma_{\xi}^{-2}$ is unique except for a sign. 

  Now, coming back to equations (\ref{meta}) and (\ref{mxi}) we have
  that 
 \begin{eqnarray*}
\Sigma_{Y \eta} &=& A (A^T \Sigma_Y^{-1} A)^{-1/2} \label{meta2}\\
\Sigma_{X \xi} &=& B  (B^T \Sigma_X^{-1} X)^{-1/2}  \label{mxi2}
\end{eqnarray*} 
and again are unique except by a sign.

Now, using again (\ref{SIGMA}), the fact that $\mu_{\eta}=\mu_{\xi} =0$ and  
$\var (E (\xi|X) = \var (E (\eta|Y)=1$ we have
 \begin{eqnarray*}
E (\eta | Y) &=&    \Sigma_{\eta Y} \Sigma_{Y}^{-1}( Y - \mu_{  Y})\\
\sigma_{\eta}^{2}& =& E( \var (\eta| Y) )+1\\
E (\xi |   X) &=&   \Sigma_{\xi X}  \Sigma_{X}^{-1}( X - \mu_{  X})\\
\sigma_{\xi}^{2}& =& E( \var (\xi | X) )+1.
\end{eqnarray*}

Replacing $\Sigma_{YX}= \Sigma_{Y \eta} \sigma_{\eta}^{-2} \sigma_{\xi \eta} \sigma_{\xi}^{-2} \Sigma_{\xi X}$ we have
\begin{eqnarray*}
 \cov \{ E (\eta |Y), E(\xi  |X)\}&=& \Sigma_{\eta Y} \Sigma_{Y}^{-1} \Sigma_{YX} \Sigma_{X}^{-1} \Sigma_{X \xi} \\
 &=&  \Sigma_{\eta Y} \Sigma_{Y}^{-1}  \Sigma_{Y \eta} \sigma_{\eta}^{-2} \sigma_{\xi \eta} \sigma_{\xi}^{-2} \Sigma_{\xi X} \Sigma_{X}^{-1} \Sigma_{X \xi} \\
 \cor \{ E (\eta |Y),  E(\xi  |X)\}&=&    (\Sigma_{\eta Y} \Sigma_{Y}^{-1}  \Sigma_{Y \eta} )^{1/2} \sigma_{\eta}^{-2} \sigma_{\xi \eta} \sigma_{\xi}^{-2}( \Sigma_{\xi X} \Sigma_{X}^{-1} \Sigma_{X \xi})^{1/2}\\
 &=&  \sigma_{\eta}^{-2} \sigma_{\xi \eta} \sigma_{\xi}^{-2}\\
 &=&  (m_\eta m_\xi)^{-1}\\
 &=& (A^T \Sigma_{Y}^{-1} A)^{1/2} (B^T \Sigma_{X}^{-1} B)^{1/2}.
  \end{eqnarray*}

Now, we will prove that $\sigma_{\xi}$ and $\sigma_{\eta}$ are not identifiable. For that, since $\sigma_{\eta} $ and $\sigma_{\xi}$ have to be greater than 1 and $\sigma_{\eta } \sigma_{\xi}$ should be constant we can change $\xi $ by $C \xi $ and $\eta $ by $C^{-1} \eta$ in such a way that $C \sigma_{\xi}>1 $ and $C^{-1} \sigma_{\eta}>1$. We take any $C$ such that
$\sigma_{\xi}^{-1}< C < \sigma_{\eta}$ and none of the other parameters change.
    From this follows that $\sigma_{\xi \eta}$ and $\beta_{X|\xi}$, $\beta_{Y|\eta}$ is not identifiable. It is left to prove that $\Sigma_{Y|\eta}$ and $\Sigma_{X|\xi}$ are not identifiable. For that we use
  $  \Sigma_X = \Sigma_{X|\xi} + \Sigma_{X \xi} \sigma_{\xi}^{-2} \Sigma_{\xi X}$.
    If $\Sigma_{X|\xi}$ is identifiable, since $ \Sigma_X$ and $ \Sigma_{X \xi} $ are identifiable, $\sigma_{\xi}$ is identifiable, which is a contradiction since we have already proven that it is not so. The same argument holds for $\Sigma_{Y|\eta}$. \end{proofprop}
      
 \begin{lemma}\label{aux}
Under the reflexive model of Section \ref{setup} and the regression constraints, 
\begin{eqnarray} 
\Sigma_X &=& \Sigma_{X|\xi} + \sigma_{\xi}^{-2} (B^T\Sigma_X^{-1} B)^{-1}  B B^T \label{aa}\\
 &=& \Sigma_{X|\xi} + \frac1{\sigma_{\xi}^{2}-1}  (B^T\Sigma_{X|\xi}^{-1} B)^{-1}  B B^T.\label{bb}\\
\Sigma_Y &=& \Sigma_{Y|\eta} + \sigma_{\eta}^{-2} (A^T\Sigma_Y^{-1} A)^{-1}  A A^T \label{cc}\\
 &=& \Sigma_{Y|\eta} + \frac1{\sigma_{\eta}^{2} -1}  (A^T\Sigma_{Y|\eta}^{-1}A)^{-1}  A A^T. \label{dd}
 \end{eqnarray}
 \end{lemma}

\proof      
      
 By the covariance formula and the fact that by Proposition \ref{one} we have $\Sigma_{X \xi} = B (B^T \Sigma_X^{-1}B)^{-1/2} $ and $\Sigma_{Y \eta} = A(A^T \Sigma_{Y}^{-1}A)^{-1/2} $ from where we get (\ref{aa}) and (\ref{cc}).
 Now,
 taking inverse and using the Woodbury inequality we have
  \begin{eqnarray*}
 B^T\Sigma_X^{-1} B &=& \sigma_{\xi}^{2}\frac{  B^T\Sigma_{X|\xi}^{-1} B  B^T\Sigma_{X}^{-1} B}{\sigma_{\xi}^{2} B^T \Sigma_X^{-1} B + B^T \Sigma_{X|\xi}^{-1} B}
 \end{eqnarray*}
 As a consequence
  \begin{eqnarray*}
 B^T\Sigma_X^{-1} B &=& B^T\Sigma_{X|\xi}^{-1} B \frac{\sigma_{\xi}^{2} -1}{\sigma_{\xi}^{2}}
 \end{eqnarray*} 
 and (\ref{bb}) follows replacing this into (\ref{aa}).
 The proof of  (\ref{dd}) follows analogously. \eop
 

\begin{prop}\label{two}
Assume the regression constraints. If $\Sigma_{Y|\eta} $ and $\Sigma_{X|\xi}$ are identifiable then $\sigma_{\xi}^{2}$, $\sigma_{\eta}^{2}$, $|\sigma_{\xi \eta}|$,  $\beta_{X|\xi}$, $ \beta_{Y|\eta}$ and $\cor (\eta, \xi) $ are identifiable, and
\begin{eqnarray}\label{formula} 
\cor (\eta, \xi) 
&=& \cor\{ E(\eta|Y),E(\xi|X)\}  \sigma_{\xi}\sigma_{\eta}\nonumber\\
\sigma_{\xi}^{2}&=& \frac{H_\xi}{H_{\xi}-1} \label{laxi}\\
\sigma_{\eta}^{2} &=& \frac{H_\eta}{H_{\eta}-1}, \label{laeta}
\end{eqnarray}
where $H_{\xi }= \Sigma_{\xi X}  \Sigma_{X|\xi}^{-1}\Sigma_{X \xi} $ and $H_{\eta } =  \Sigma_{\eta Y} \Sigma_{Y|\eta}^{-1} \Sigma_{Y \eta}$.
Moreover,  $\Sigma_{Y|\eta} $ and $\Sigma_{X|\xi}$ are identifiable if and only if $\sigma_{\xi}^{2}$, $\sigma_{\eta}^{2}$ are so.

\end{prop}
\proof
 
By (\ref{aa})-(\ref{dd}) and using Proposition  \ref{one}  we have that $\sigma_{\xi}$ and $\sigma_{\eta}$ are identifiable and as a consequence the rest of the parameters are identifiable. 
Now, to prove (\ref{laxi}) let us use the formula
\begin{eqnarray*}
\Sigma_X &=& \Sigma_{X|\xi} + \Sigma_{X\xi} \sigma_{\xi}^{-2} \Sigma_{\xi X}.
\end{eqnarray*}
Using Woodbury inequality
\begin{eqnarray}\label{mmm}
\Sigma_X^{-1} &=& \Sigma_{X|\xi}^{-1} - \Sigma_{X|\xi}^{-1} \Sigma_{X\xi}(\sigma_{\xi}^{2}+ \Sigma_{\xi X} \Sigma_{X\xi}^{-1} \Sigma_{X \xi})^{-1} \Sigma_{\xi X} \Sigma_{X|\xi}^{-1}.
\end{eqnarray}
Using the fact that $\Sigma_{\xi X} \Sigma_X^{-1}  \Sigma_{X \xi}=1 $ proven in Proposition \ref{one} and multiplying to the left and to the right of (\ref{mmm}) by $\Sigma_{\xi X}$ and $\Sigma_{X \xi}$  we have
\begin{eqnarray*}
1 &=&\Sigma_{\xi X} \Sigma_{X|\xi}^{-1} \Sigma_{X\xi}-\Sigma_{\xi X} \Sigma_{X|\xi}^{-1} \Sigma_{X\xi} (\sigma_{\xi}^{2}+ \Sigma_{\xi X} \Sigma_{X|\xi} ^{-1} \Sigma_{X \xi})^{-1} \Sigma_{\xi X} \Sigma_{X|\xi}^{-1}\Sigma_{X\xi}\\
& =& H_{\xi} - H_{\xi} ( \sigma_{\xi}^{2} + H_\xi)^{-1} H_{\xi}\\
&=& \frac{H_\xi   \sigma_{\xi}^{2}}{\sigma_{\xi}^{2}+ H_{\xi}}
\end{eqnarray*}
from where we get (\ref{laxi}). Analogously we get (\ref{laeta}). \eop 

\begin{proofprop}{prop:three1}
Identification of $\Sigma_{X|\xi} $ means that if
we have two different matrices $\Sigma_{X} $ and $ \tilde{\Sigma}_{X}$
 satisfying (\ref{a}) then we should conclude that $\Sigma_{X|\xi} =\tilde{\Sigma}_{X|\xi}$ and that $\sigma_{\xi}^{2} = \sigmat_{\xi}^{2}$. The same logic applies to $\Sigma_{Y|\eta}$ and $\sigma_{\eta}^{2}$. More specifically, from (\ref{a}) the equality
  \begin{eqnarray}
  \Sigma_{X|\xi}  + \sigma_{\xi}^{-2} (B^T\Sigma_X^{-1} B)^{-1}BB^T &=&
  \tilde{{\Sigma}}_{X|\xi}  + \tilde{\sigma}_{\xi}^{-2} (B^T\Sigma_X^{-1} B)^{-1} BB^T \label{e}
  \end{eqnarray}
must  imply that $\Sigma_{X|\xi}=\tilde{\Sigma}_{X|\xi}$ and $\sigma_{\xi}^{2} = \tilde{\sigma}_{\xi}^{2}$.  If $\sigma_{\xi}^{2}$ is identifiable, so $\sigma_{\xi}^{2} = \tilde{\sigma}_{\xi}^{2}$, then (\ref{e}) implies $\Sigma_{X|\xi}=\tilde{\Sigma}_{X|\xi}$ since $(B^T\Sigma_X^{-1} B)^{-1}BB^T$ is identifiable.  Similarly, if $\Sigma_{X|\xi}$ is identifiable, so $\Sigma_{X|\xi}=\tilde{\Sigma}_{X|\xi}$, then (\ref{e}) implies that $\sigma_{\xi}^{2} = \tilde{\sigma}_{\xi}^{2}$ and thus that $\sigma_{\xi}^{2}$ is identifiable.  

Now, assume that elements $(i,j)$ and $(j,i)$ of $\Sigma_{X|\xi} $ and $ \tilde{{\Sigma}}_{X|\xi}$ are known to be 0 and let $e_{k}$ denote the $p \times 1$ vector with a 1 in position $k$ and $0$'s elsewhere.  Then multiplying (\ref{e}) on the left by $e_{i}$ and on the right by $e_{j}$ gives 
\[
 \sigma_{\xi}^{-2} (B^T\Sigma_X^{-1} B)^{-1}(B)_{i}(B)_{j} =
  \tilde{\sigma}_{\xi}^{-2} (B^T\Sigma_X^{-1} B)^{-1} (B)_{i}(B)_{j}
\]
Since $(B^T\Sigma_X^{-1} B)^{-1}BB^T$ is identifiable and $(B)_{i}(B)_{j} \neq 0$, this implies that $\sigma_{\xi}^{-2} = \tilde{\sigma}_{\xi}^{-2}$ and thus $\sigma_{\xi}^{2}$ is identifiable, which implies that $\Sigma_{X|\xi}$ is identifiable.
\end{proofprop}

\begin{prooflem}{sem-conditions}
If we have $\sigma_{\xi}^{2} = \sigma_{\eta}^{2} =1$
and we define $\tilde{\xi} = c_{\xi} \xi $, $\tilde{\eta} = c_{\eta} \xi $  with $c_{\xi} =   [\var \{E(\xi|X) \}]^{-1/2}$ (we can defined because $\var E(\xi|X)$ is identifible) we have that $\cor (\tilde{\eta},\tilde{\xi})
= \cor (\xi, \eta)$.
  Now, if $\var\{E(\eta|Y)\} = \var \{E(\xi|X)\}=1$ and $\Sigma_{X|\xi}$ and $\Sigma_{Y|\eta}$ are identifiable we can identify $\sigma_{\xi}$ and $\sigma_{\eta}$ and we could define
 $\tilde{\xi} = c_{\xi} \xi $, $\tilde{\eta} = c_{\eta} \xi $  with $c_{\xi} =   \{\sigma_{\xi}\}^{-1/2}$ and
 $c_{\eta} =   \{\sigma_{\eta} \}^{-1/2}$ since $c_{\xi}$ and $c_{\eta}$ are identifiable.
\end{prooflem}

\begin{prop} \label{computer}
The SEM models under the marginal and regression constraints are
\begin{itemize}
\item Marginal constraints, \[ \left( \begin{array}{c}X \\ Y  \end{array} \right)
\sim  \left( \begin{array}{c}\mu_{X} \\  \mu_{Y}  \end{array} \right) + \epsilon_{X,Y}\]
where  
\begin{equation}\label{SIGMA2} \var ( \epsilon_{X,Y }) := \Sigma_{X,Y} = \left(\begin{array}{cc} D_{X|\xi} +  c^2 B B^T &B A^T  \\
A B^T &  D_{Y|\eta} +d^2 A A^T 
 \end{array}\right). \end{equation}
and $c^2  d^2  \cor^2(\eta,\xi) =1$.
\item Regression constraints
\[ \left( \begin{array}{c}X \\ Y  \end{array} \right)
\sim  \left( \begin{array}{c}\mu_{X} \\  \mu_{Y}  \end{array} \right) + \epsilon_{X,Y}\]
From Lemma \ref{RRR} and (\ref{bb}),
\[
\Sigma_{X,Y} = \left(\begin{array}{cc} D_{X|\xi} +  (B^T D_{X|\xi}^{-1} B )^{-1} BB^T/(\sigma_{\xi}^{2}-1)  &B A^T  \\
A B^T &  D_{Y|\eta} + (A^T D_{Y|\eta}^{-1} A )^{-1} A A^T /(\sigma_{\eta}^{2}-1)
 \end{array}\right).
 \]
 \end{itemize}
 where $\cor^2 (\eta,\xi) (\sigma_{\eta}^{2}-1)^{-1}  (A^T D_{Y|\eta}^{-1} A )^{-1}
(\sigma_{\xi}^{2}-1)^{-1}  (B^T D_{X|\xi}^{-1} B )^{-1}=1$.
 \end{prop}

\proof
For (\ref{SIGMA2}) we only need to prove $\Sigma_X = D_{X\mid \xi} + BB^T c^2$ and $\Sigma_Y= D_{Y\mid \eta} + AA^T d^2$ with $c^2d^2 \cor(\eta,\xi) = 1$.

Now, $\Sigma_X = D_{X\mid \xi} + \Sigma_{X\xi} \Sigma_{\xi X}$ and
 $\Sigma_Y = D_{Y\mid \eta} + \Sigma_{Y\eta} \Sigma_{\eta Y}$. And
 since $\Sigma_{YX} = A B^T = \Sigma_{Y \eta} \cor (\eta, \xi) \Sigma_{\xi X}$ we have
 that $\Sigma_{Y \eta }  = A d$ and $\Sigma_{X \xi} = B c$ with
$d  c \times \cor (\eta,\xi) =1$ from what follows (\ref{SIGMA2}).  Then the conclusion for the regression constraints follows from the proof of Lemma~\ref{aux}. \eop

From this we conclude that the marginal and regression constraints lead to the same estimators.

\section{Bias} \label{sec:bias}

To justify the result that 
\begin{eqnarray*}
\cor\{E(\xit \mid X), E(\etat \mid Y)\} & = & [\var\{E(\xi \mid X)\}  \var\{E(\eta \mid Y)\} ]^{-1/2} \cov\{E(\xi \mid X), E(\eta \mid Y)\} \\
& = & [\var\{E(\xi \mid X)\}  \var\{E(\eta \mid Y)\} ]^{1/2}\cor(\xi, \eta)
\end{eqnarray*}
we focus on the second equality since the first is straightforward. Now,
\begin{eqnarray*}
 \cov\{E(\xi \mid X), E(\eta \mid Y)\}  & = & \cov(\Sigma_{\xi X} \Sigma_{X}^{-1}X, \Sigma_{\eta Y} \Sigma_{Y}^{-1}Y)\\
 & = & \Sigma_{\xi X} \Sigma_{X}^{-1}\Sigma_{X Y} \Sigma_{Y}^{-1}\Sigma_{Y \eta}.
 \end{eqnarray*}
 From (\ref{SIGMA}), $\Sigma_{XY} = \beta_{X|\xi}\beta_{Y|\eta}\sigma_{\xi \eta}$.  Because $\sigma_{\xi}^{2} = \sigma_{\eta}^{2}=1$ this can be re-expressed as 
 $\Sigma_{X Y } = \Sigma_{X \xi} \Sigma_{\eta Y}\sigma_{\xi \eta} = \Sigma_{X \xi} \Sigma_{ \eta Y}\cor(\xi, \eta)$ and so
\begin{eqnarray*}
 \cov\{E(\xi \mid X), E(\eta \mid Y)\}  
 & = & \Sigma_{\xi X} \Sigma_{X}^{-1} \Sigma_{X \xi} \Sigma_{ \eta Y}\Sigma_{Y}^{-1}\Sigma_{Y \eta}\cor(\xi, \eta) \\
 & = &  [\var\{E(\xi \mid X)\}  \var\{E(\eta \mid Y)\} ]\cor(\xi, \eta).
 \end{eqnarray*}
 Multiplying both sides by $ [\var\{E(\xi \mid X)\}  \var\{E(\eta \mid Y)\} ]^{-1/2}$ completes the justification.
 
To perhaps aid intuition, we can see the result in another way by using the Woodbury identity to invert $\Sigma_{X} = (\Sigma_{X \xi} \Sigma_{\xi X} + \Sigma_{X|\xi})$ we have 
 \begin{eqnarray*}
\Sigma_{\xi X} \Sigma_{X}^{-1} \Sigma_{X \xi}& = &\Sigma_{\xi X} (\Sigma_{X \xi} \Sigma_{\xi X} + \Sigma_{X|\xi})^{-1}\Sigma_{X \xi} \\
 & = & \frac{\Sigma_{\xi X} \Sigma_{X|\xi}^{-1}\Sigma_{X \xi} }{1+\Sigma_{\xi X} \Sigma_{X|\xi}^{-1}\Sigma_{X \xi} }.
 \end{eqnarray*}
 Similarly,
  \begin{eqnarray*}
\Sigma_{\eta Y} \Sigma_{Y}^{-1} \Sigma_{Y \eta}
 & = & \frac{\Sigma_{\eta Y} \Sigma_{Y|\eta}^{-1}\Sigma_{Y \eta} }{1+\Sigma_{\eta Y} \Sigma_{Y|\eta}^{-1}\Sigma_{Y \eta} },
\end{eqnarray*}
and thus
\[
\cor\{E(\xit \mid X), E(\etat \mid Y)\} =
\left[ \frac{\Sigma_{\xi X} \Sigma_{X|\xi}^{-1}\Sigma_{X \xi} }{1+\Sigma_{\xi X} \Sigma_{X|\xi}^{-1}\Sigma_{X \xi} }  \frac{\Sigma_{\eta Y} \Sigma_{Y|\eta}^{-1}\Sigma_{Y \eta} }{1+\Sigma_{\eta Y} \Sigma_{Y|\eta}^{-1}\Sigma_{Y \eta} } \right]^{1/2}\cor(\xi, \eta).
\]
This form again shows that $\cor\{E(\xit \mid X), E(\etat \mid Y)\} \leq \cor(\xi, \eta)$ and provides a more detailed expression of their ratio.

\section{Additional examples}\label{sec:addexamples}

\begin{figure}[ht!]
\begin{center}

\includegraphics[width=0.6\textwidth]{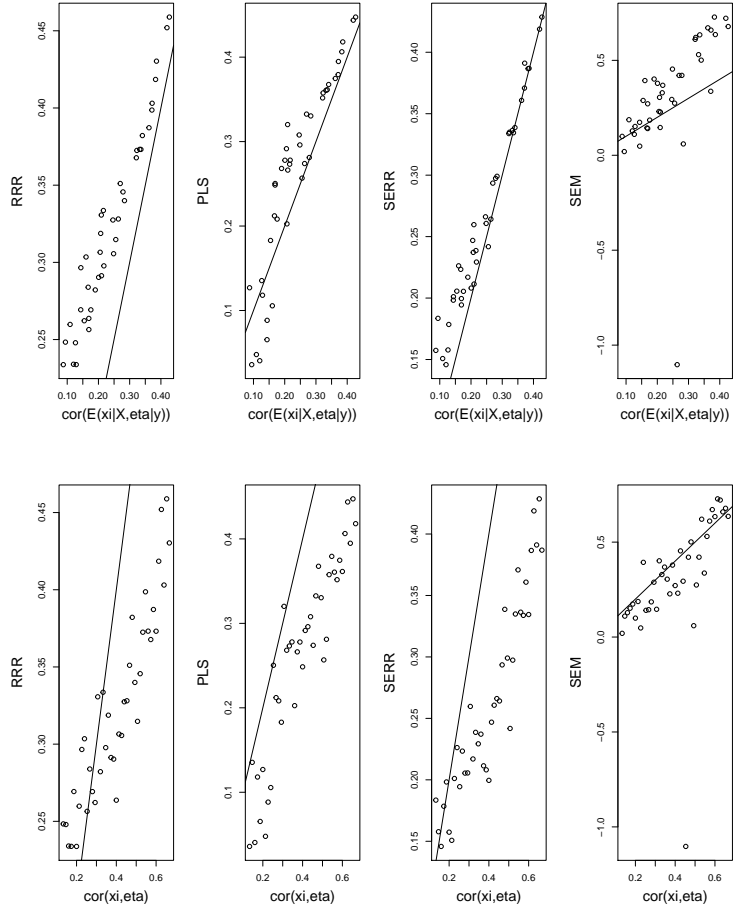}  
 \caption{$L=  (0.58 ,0.98, 0)$, $\Sigma_{X|\xi}=\Sigma_{Y|\eta}=I_3$, N=100.}\label{fig2}
\end{center}
\vspace{-0.4cm}
\end{figure}

Figure~\ref{fig2} shows simulations results constructed as for Figure~\ref{fig1}, except  $L=(0.58 ,0.98, 0)^{T}$ to reflect a weak signal, so $\cor\{E(\eta \mid Y), E(\xi \mid X)\} = 0.56 \cor(\eta,\xi)$.  In this case none of methods do very well.  SEM seems prone outlying results, underestimates $\cor\{E(\eta \mid Y), E(\xi \mid X)\}$, as expected, and underestimates $\cov(\xi,\eta)$.  Overall, SERR seems to do the best, although it overestimates its target for small covariances because then accurate estimation of the weights is difficult.

\begin{figure}[ht!]

\begin{center}

\includegraphics[width=0.6\textwidth]{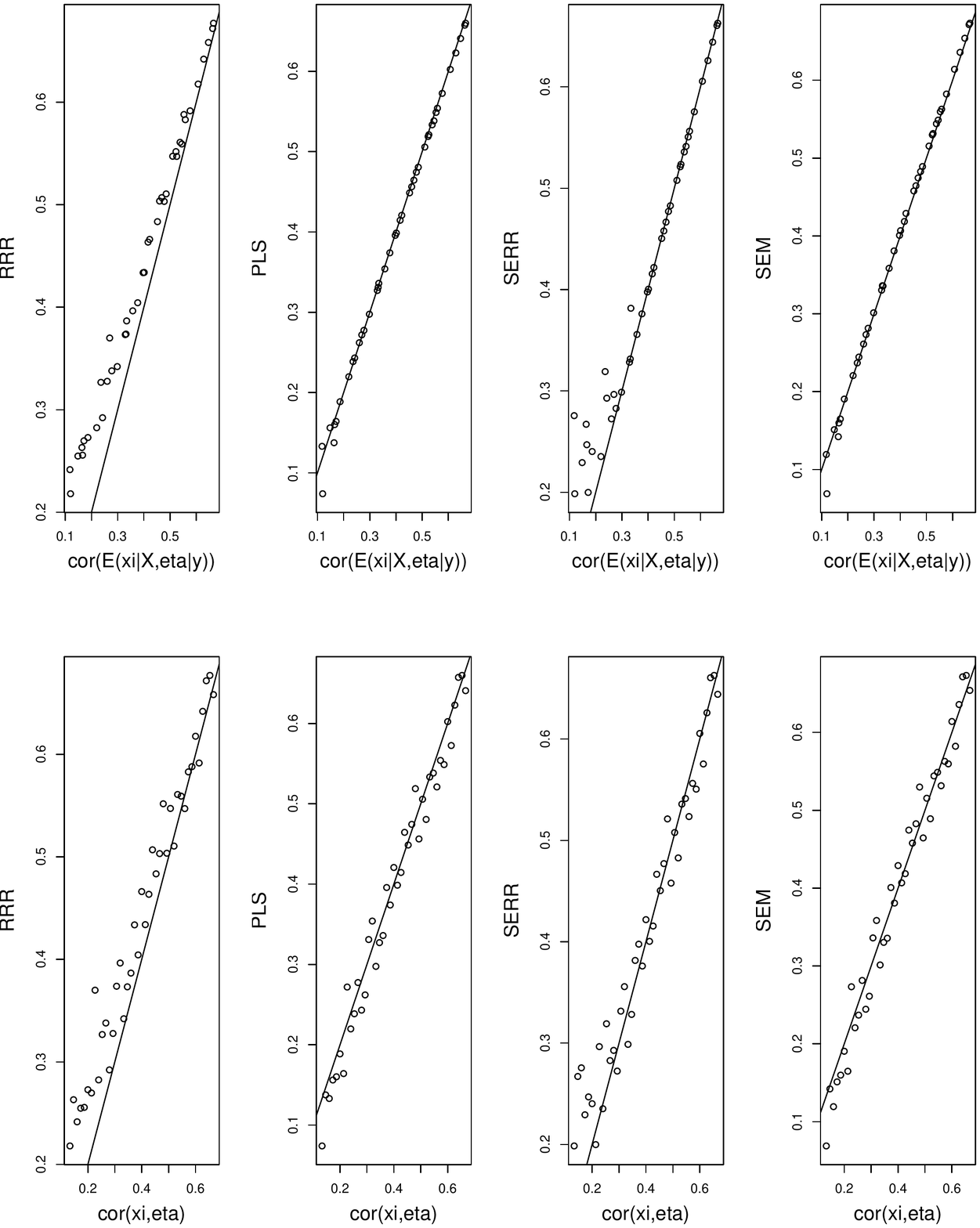} 
\caption{$L=  (4,4,4)^{T}$, $\Sigma_{X|\xi}=\Sigma_{Y|\eta}=I_3$, $N=100$}\label{fig4}
\end{center}
\vspace{-0.4cm}
\end{figure}

Figure~\ref{fig4} shows simulations results constructed as for Figure~\ref{fig2}, except now to induce a large overall signal we took $L=(4,4,4)^{T}$, so $\cor\{E(\eta \mid Y), E(\xi \mid X)\} = (48/49) \cor(\eta,\xi)$.  The performance of all methods improved over Figure~\ref{fig2}, as expected. SERR still has a  tendency to overestimate for small correlations, while moment-based PLS did better in such settings.  Although SERR is a likelihood-based method, its asymptotic properties do not necessarily hold for small samples.  In particular, these results support other studies showing empirically that moment-based PLS can do better than likelihood-based PLS in small samples.  The meaning of ``small'' in this context depends on the size of the signal.  For Figures~\ref{fig2} and \ref{fig4} the signal size depends on both $L$ and $\cor(\xi, \eta)$.

    \section*{References}

\bibliography{ref}

\begin{thebibliography}{}

\bibitem[\protect\citeauthoryear{Akter, Wamba, and Dewan}{Akter
  et~al.}{2017}]{Akter2017}
Akter, S., S.~F. Wamba, and S.~Dewan (2017).
\newblock Why pls-sem is suitable for complex modelling? an empirical
  illustration in big data analytics quality.
\newblock {\em Production Planning \& Control\/}~{\em 28\/}(11-12), 1011--1021.

\bibitem[\protect\citeauthoryear{Anderson}{Anderson}{1951}]{Anderson1951}
Anderson, T.~W. (1951).
\newblock Estimating linear restrictions on regression coefficients for
  multivariate normal distributions.
\newblock {\em Ann. Math. Statist.\/}~{\em 22\/}(3), 327--351.

\bibitem[\protect\citeauthoryear{Bollen}{Bollen}{1989}]{Bollen1989}
Bollen, K.~A. (1989).
\newblock {\em Structural Equations with Latent Variables}.
\newblock New York: Wiley.

\bibitem[\protect\citeauthoryear{Chun and Kele\c{s}}{Chun and
  Kele\c{s}}{2010}]{Chun2010}
Chun, H. and S.~Kele\c{s} (2010).
\newblock Sparse partial least squares regression for simultaneous dimension
  reduction and predictor selection.
\newblock {\em Journal of the Royal Statistical Society B\/}~{\em 72\/}(1),
  3--25.

\bibitem[\protect\citeauthoryear{Cook and Forzani}{Cook and
  Forzani}{2017}]{CookForzani2017PLS}
Cook, R.~D. and L.~Forzani (2017).
\newblock Big data and partial least squares prediction.
\newblock {\em The Canadian Journal of Statistics/La Revue Canadienne de
  Statistique\/}~{\em to appear}.

\bibitem[\protect\citeauthoryear{Cook and Forzani}{Cook and
  Forzani}{2018}]{CookForzani2018}
Cook, R.~D. and L.~Forzani (2018).
\newblock Partial least squares prediction in high-dimensional regression.
\newblock {\em Annals of Statistics\/}~{\em to appear}.

\bibitem[\protect\citeauthoryear{Cook, Forzani, and Rothman}{Cook
  et~al.}{2012}]{Cook2012estimating}
Cook, R.~D., L.~Forzani, and A.~J. Rothman (2012).
\newblock Estimating sufficient reductions of the predictors in abundant
  high-dimensional regressions.
\newblock {\em The Annals of Statistics\/}~{\em 40\/}(1), 353--384.

\bibitem[\protect\citeauthoryear{Cook, Forzani, and Rothman}{Cook
  et~al.}{2013}]{Cook2013}
Cook, R.~D., L.~Forzani, and A.~J. Rothman (2013).
\newblock Prediction in abundant high-dimensional linear regression.
\newblock {\em Electronic Journal of Statistics\/}~{\em 7}, 3059--3088.

\bibitem[\protect\citeauthoryear{Cook, Forzani, and Zhang}{Cook
  et~al.}{2015}]{Cook:2015ab}
Cook, R.~D., L.~Forzani, and X.~Zhang (2015).
\newblock Envelopes and reduced-rank regression.
\newblock {\em Biometrika\/}~{\em 102\/}(2), 439--456.

\bibitem[\protect\citeauthoryear{Cook, Helland, and Su}{Cook
  et~al.}{2013}]{CookHellandSu2013}
Cook, R.~D., I.~S. Helland, and Z.~Su (2013).
\newblock Envelopes and partial least squares regression.
\newblock {\em Journal of the Royal Statistical Society: Series B (Statistical
  Methodology)\/}~{\em 75\/}(5), 851--877.

\bibitem[\protect\citeauthoryear{Cook, Li, and Chiaromonte}{Cook
  et~al.}{2007}]{Cook2007}
Cook, R.~D., B.~Li, and F.~Chiaromonte (2007).
\newblock Dimension reduction in regression without matrix inversion.
\newblock {\em Biometrika\/}~{\em 94\/}(3), 569--584.

\bibitem[\protect\citeauthoryear{Cook, Li, and Chiaromonte}{Cook
  et~al.}{2010}]{Cook2010}
Cook, R.~D., B.~Li, and F.~Chiaromonte (2010).
\newblock Envelope models for parsimonious and efficient multivariate linear
  regression (with discussion).
\newblock {\em Statist. Sci.\/}~{\em 20}, 927--1010.

\bibitem[\protect\citeauthoryear{Cook and Zhang}{Cook and
  Zhang}{2015a}]{CookZhang2015foundation}
Cook, R.~D. and X.~Zhang (2015a).
\newblock Foundations for envelope models and methods.
\newblock {\em Journal of the American Statistical Association\/}~{\em
  110\/}(510), 599--611.

\bibitem[\protect\citeauthoryear{Cook and Zhang}{Cook and
  Zhang}{2015b}]{CookZhang2015}
Cook, R.~D. and X.~Zhang (2015b).
\newblock Simultaneous envelopes for multivariate linear regression.
\newblock {\em Technometrics\/}~{\em 57\/}(1), 11--25.

\bibitem[\protect\citeauthoryear{de~Jong}{de~Jong}{1993}]{deJong1993}
de~Jong, S. (1993).
\newblock Simpls: An alternative approach to partial least squares regression.
\newblock {\em Chemometrics and Intelligent Laboratory Systems\/}~{\em
  18\/}(3), 251--263.

\bibitem[\protect\citeauthoryear{Frank and Frideman}{Frank and
  Frideman}{1993}]{Frank1993}
Frank, I.~E. and J.~H. Frideman (1993).
\newblock A statistical view of some chemometrics regression tools.
\newblock {\em Technometrics\/}~{\em 35\/}(2), 102--246.

\bibitem[\protect\citeauthoryear{Galadi}{Galadi}{1988}]{Geladi1988}
Galadi, P. (1988).
\newblock Notes on the history and nature of partial least squares (pls)
  modeling.
\newblock {\em Journal of Chemometrics\/}~{\em 2\/}(4), 231--246.

\bibitem[\protect\citeauthoryear{Guide and Ketokivi}{Guide and
  Ketokivi}{2015}]{Guide2015}
Guide, J.~B. and M.~Ketokivi (2015, July).
\newblock Notes from the editors: Redefining some methodological criteria for
  the journal.
\newblock {\em Journal of Operations Management\/}~{\em 37}, v--viii.

\bibitem[\protect\citeauthoryear{Helland}{Helland}{1990}]{Helland1990}
Helland, I.~S. (1990).
\newblock Partial least squares regression and statistical models.
\newblock {\em Scandinavian Journal of Statistics\/}~{\em 17\/}(2), 97--114.

\bibitem[\protect\citeauthoryear{Helland}{Helland}{1992}]{Helland1992}
Helland, I.~S. (1992).
\newblock Maximum likelihood regression on relevant components.
\newblock {\em Journal of the Royal Statistical Society B\/}~{\em 54\/}(2),
  637--647.

\bibitem[\protect\citeauthoryear{Henseler, Dijkstra, Sarstedt, Ringle,
  Diamantopoulos, Straub, J.~Ketchen~Jr, F.~Hair, Hult, and Calantone}{Henseler
  et~al.}{2014}]{Henseler2014}
Henseler, J., T.~Dijkstra, M.~Sarstedt, C.~Ringle, A.~Diamantopoulos,
  D.~Straub, D.~J.~Ketchen~Jr, J.~F.~Hair, T.~Hult, and R.~Calantone (2014).
\newblock Common beliefs and reality about pls: Comments on r\"{o}nkk\"{o} \&
  evermann (2013).
\newblock {\em Organizational Research Methods\/}~{\em 17\/}(2), 182--209.

\bibitem[\protect\citeauthoryear{Hui and Wold}{Hui and Wold}{1982}]{Hui1982}
Hui, B.~S. and H.~Wold (1982).
\newblock Consistency and consistency at large of partial least squares
  estimates.
\newblock In K.~G. J\"{o}reskog and H.~Wold (Eds.), {\em Systems under indirect
  observation, part II}. Amsterdam: North Holland.

\bibitem[\protect\citeauthoryear{Izenman}{Izenman}{1975}]{Izenman:1975aa}
Izenman, A.~J. (1975).
\newblock Reduced-rank regression for the multivariate linear model.
\newblock {\em Journal of Multivariate Analysis\/}~{\em 5\/}(2), 248--264.

\bibitem[\protect\citeauthoryear{J\"{o}reskog}{J\"{o}reskog}{1970}]{joreskog1970}
J\"{o}reskog, K.~G. (1970, September).
\newblock A general method for estimating a linear structural equation model.
\newblock Technical Report RB-70-54, Educational Testing Service, Princeton,
  New Jersey.

\bibitem[\protect\citeauthoryear{Lohm{\"o}ller}{Lohm{\"o}ller}{1989}]{LohmllerJ.-B.1989}
Lohm{\"o}ller, J.-B. (1989).
\newblock {\em Latent Variable Path Modeling with Partial Least Squares}.
\newblock Springer, New York.

\bibitem[\protect\citeauthoryear{McIntosh, Edwards, and Antonakis}{McIntosh
  et~al.}{2014}]{Mcintosh2014}
McIntosh, C.~N., J.~R. Edwards, and J.~Antonakis (2014).
\newblock Reflections on partial least squares path modeling.
\newblock {\em Organizational Research Methods\/}~{\em 17\/}(2), 210--251.

\bibitem[\protect\citeauthoryear{Reinsel and Velu}{Reinsel and
  Velu}{1998}]{ransey}
Reinsel, G. and R.~Velu (1998).
\newblock {\em Multivariate Reduced-Rank Regression - Theory and Applications}.
\newblock Lecture Notes in Statistics. Springer, New York.

\bibitem[\protect\citeauthoryear{Rigdon}{Rigdon}{2012}]{Rigdon2012}
Rigdon (2012).
\newblock Rethinking partial least squares path modeling: In praise of simple
  methods.
\newblock {\em Long Range Planning\/}~{\em 45}, 341--354.

\bibitem[\protect\citeauthoryear{R{\"o}nkk{\"o} and Evermann}{R{\"o}nkk{\"o}
  and Evermann}{2013}]{Ronkko2013}
R{\"o}nkk{\"o}, M. and J.~Evermann (2013, 2017/08/17).
\newblock A critical examination of common beliefs about partial least squares
  path modeling.
\newblock {\em Organizational Research Methods\/}~{\em 16\/}(3), 425--448.

\bibitem[\protect\citeauthoryear{R\"{o}nkk\"{o}, McIntosh, Antonakis, and
  Edwards}{R\"{o}nkk\"{o} et~al.}{2016}]{Ronkko2016}
R\"{o}nkk\"{o}, M., C.~N. McIntosh, J.~Antonakis, and J.~R. Edwards (2016).
\newblock Partial least squares path modeling: Time for some serious second
  thoughts.
\newblock {\em Journal of Operations Management\/}~{\em 47--48\/}(Supplement
  C), 9--27.

\bibitem[\protect\citeauthoryear{Sarstedt, Hair, Ringle, Thiele, and
  Gudergan}{Sarstedt et~al.}{2016}]{Sarstedt2016}
Sarstedt, M., J.~F. Hair, C.~M. Ringle, K.~O. Thiele, and S.~P. Gudergan
  (2016).
\newblock Estimation issues with pls and cbsem: Where the bias lies!
\newblock {\em Journal of Business Research\/}~{\em 69\/}(10), 3998 -- 4010.

\bibitem[\protect\citeauthoryear{Tenenhaus and Vinzi}{Tenenhaus and
  Vinzi}{2005}]{Tenenhaus2005}
Tenenhaus, M. and E.~V. Vinzi (2005).
\newblock Pls regression, pls path modeling, and generalized procrustean
  analysis: a combined approach for multiblock analysis.
\newblock {\em Journal of Chemometrics\/}~{\em 19\/}(3), 145--153.

\bibitem[\protect\citeauthoryear{Tipping and Bishop}{Tipping and
  Bishop}{1999}]{TippingBishop1999}
Tipping, M.~E. and C.~M. Bishop (1999).
\newblock Probabilistic principal component analysis.
\newblock {\em Journal of the Royal Statistical Society B\/}~{\em 61\/}(3),
  611--622.

\bibitem[\protect\citeauthoryear{Vinzi, Trinchera, and Amato}{Vinzi
  et~al.}{2010}]{Vinzi2010}
Vinzi, E.~V., L.~Trinchera, and S.~Amato (2010).
\newblock Pls path modeling: From foundations to recent developments and open
  issues for model assessment and improvement.
\newblock In E.~V. Vinzi, W.~W. Chin, J.~Henseler, and H.~Wang (Eds.), {\em
  Handbook of Partial Least Squares}, Chapter~2, pp.\  47--82. Berlin:
  Springer-Verlag.

\bibitem[\protect\citeauthoryear{Wold, Martens, and Wold}{Wold
  et~al.}{1983}]{Wold1983}
Wold, S., H.~Martens, and H.~Wold (1983).
\newblock The multivariate calibration problem in chemistry solved by the pls
  method.
\newblock In A.~Ruhe and B.~K\aa~gstr\"{o}m (Eds.), {\em Proceedings of the
  Conference on Matrix Pencils, Lecture Notes in Mathematics}, Volume 973, pp.\
   286--293. Heidelberg: Springer Verlag.

\end{thebibliography}

\end{document}